\documentclass[manuscript]{aastex63} 

\usepackage{float} 
\usepackage{siunitx}
\usepackage{xcolor}
\usepackage{cancel}
\usepackage{amsmath}

\shorttitle{ Mini-EUSO UV terrestrial emissions and maps}

\begin{document}

\title{Observation of night-time emissions of the Earth in the near UV range from the  International Space Station with the Mini-EUSO detector}

\author[0000-0001-6067-5104]{M Casolino}
\affiliation{ INFN, Sezione di Roma Tor Vergata, Roma, Italy}
\affiliation{ Universit\`a degli Studi di Roma Tor Vergata, Dipartimento di Fisica, Roma, Italy}
\affiliation{ RIKEN, Wako, Japan}
\author{D Barghini}
\affiliation{ INFN, Sezione di Torino, Torino, Italy}
\affiliation{ Dipartimento di Fisica, Universit\`a di Torino, Italy}
\affiliation{ INAF, Osservatorio astrofisico di Torino, Torino, Italy}
\author[0000-0001-5686-4711]{M Battisti}
\affiliation{ INFN, Sezione di Torino, Torino, Italy}
\author{C Blaksley}
\affiliation{ RIKEN, Wako, Japan}
\author[0000-0002-8565-6409]{A Belov}
\affiliation{ Faculty of Physics, M.V. Lomonosov Moscow State University, Moscow, Russia}
\affiliation{ Skobeltsyn Institute of Nuclear Physics, Lomonosov Moscow State University, Moscow, Russia}
\author[0000-0003-1069-1397]{M Bertaina}
\affiliation{ INFN, Sezione di Torino, Torino, Italy}
\affiliation{ Dipartimento di Fisica, Universit\`a di Torino, Italy}
\author{M Bianciotto}
\affiliation{ Dipartimento di Fisica, Universit\`a di Torino, Italy}
\author[0000-0002-1606-125X]{F Bisconti}
\affiliation{ INFN, Sezione di Roma Tor Vergata, Roma, Italy}
\affiliation{ INFN, Sezione di Torino, Torino, Italy}
\author{S Blin} 
\affiliation{ APC, Univ Paris Diderot, CNRS/IN2P3, CEA/Irfu, Obs de Paris, Sorbonne Paris Cit\'e, France}\author[0000-0001-5456-3894]{K Bolmgren}
\affiliation{ KTH Royal Institute of Technology, Stockholm, Sweden}
\author[0000-0003-0711-8768]{G Cambi\`e}
\affiliation{ INFN, Sezione di Roma Tor Vergata, Roma, Italy}
\affiliation{ Universit\`a degli Studi di Roma Tor Vergata, Dipartimento di Fisica, Roma, Italy}
\author[0000-0002-1153-2139]{F Capel}
\affiliation{ Max Planck Institute for Physics, Munich, Germany}
\affiliation{ KTH Royal Institute of Technology, Stockholm, Sweden}
\author{I Churilo}
\affiliation{ S.P. Korolev Rocket and Space Corporation Energia, Korolev, Moscow area, Russia}
\author[0000-0001-9457-1338]{M Crisconio}
\affiliation{ ASI, Italian Space Agency, Rome. Italy}
\author{C De La Taille} 
\affiliation{Omega, Ecole Polytechnique, CNRS/IN2P3, Palaiseau, France}
\author{T Ebisuzaki}
\affiliation{ RIKEN, Wako, Japan}
\author[0000-0003-3849-2955]{J Eser}
\affiliation{ Department of Astronomy and Astrophysics, The University of Chicago, IL, USA}
\author{F Fenu} 
\affiliation{ KIT, Germany}
\author{M A Franceschi} 
\affiliation{ INFN-LNF, Frascati, Italy}
\author[0000-0002-0406-0962]{C Fuglesang} 
\affiliation{ KTH Royal Institute of Technology, Stockholm, Sweden}
\author[0000-0001-7674-2172]{A Golzio}
\affiliation{ INFN, Sezione di Torino, Torino, Italy}
\affiliation{ Dipartimento di Fisica, Universit\`a di Torino, Italy}
\author{P Gorodetzky} 
\affiliation{ APC, Univ Paris Diderot, CNRS/IN2P3, CEA/Irfu, Obs de Paris, Sorbonne Paris Cit\'e, France}
\author{H Kasuga}
\affiliation{ RIKEN, Wako, Japan}
\author{F Kajino} 
\affiliation{ Konan University, Kobe, Japan}
\author[0000-0001-9815-6123]{P Klimov}
\affiliation{ Faculty of Physics, M.V. Lomonosov Moscow State University, Moscow, Russia}
\affiliation{ Skobeltsyn Institute of Nuclear Physics, Lomonosov Moscow State University, Moscow, Russia}
\author{V Kuznetsov} 
\affiliation{ S.P. Korolev Rocket and Space Corporation Energia, Korolev, Moscow area, Russia}
\author{M Manfrin}
\affiliation{ INFN, Sezione di Torino, Torino, Italy}
\affiliation{ Dipartimento di Fisica, Universit\`a di Torino, Italy}
\author[0000-0002-3180-1228]{L Marcelli}
\affiliation{ INFN, Sezione di Roma Tor Vergata, Roma, Italy}
\author{G Mascetti}
\affiliation{ ASI, Italian Space Agency, Rome. Italy}
\author[0000-0001-7604-5473]{W Marsza{\l}}
\affiliation{ National Centre for Nuclear Research, Lodz, Poland}
\author{H Miyamoto}
\affiliation{ INFN, Sezione di Torino, Torino, Italy}
\author{A Murashov}
\affiliation{ Faculty of Physics, M.V. Lomonosov Moscow State University, Moscow, Russia}
\affiliation{ Skobeltsyn Institute of Nuclear Physics, Lomonosov Moscow State University, Moscow, Russia}
\author{T Napolitano} 
\affiliation{ INFN-LNF, Frascati, Italy}
\author{H Ohmori}
\affiliation{ RIKEN, Wako, Japan}
\author{A Olinto} 
\affiliation{ Department of Astronomy and Astrophysics, The University of Chicago, IL, USA}
\author{E Parizot} 
\affiliation{ APC, Univ Paris Diderot, CNRS/IN2P3, CEA/Irfu, Obs de Paris, Sorbonne Paris Cit\'e, France}
\author[0000-0002-7986-3321]{P Picozza} 
\affiliation{ INFN, Sezione di Roma Tor Vergata, Roma, Italy}
\affiliation{ Universit\`a degli Studi di Roma Tor Vergata, Dipartimento di Fisica, Roma, Italy}
\author[0000-0002-1523-3398]{L W Piotrowski} 
\affiliation{ Faculty of Physics, University of Warsaw, Warsaw, Poland}
\author{Z Plebaniak} 
\affiliation{ INFN, Sezione di Torino, Torino, Italy}
\affiliation{ Dipartimento di Fisica, Universit\`a di Torino, Italy}
\affiliation{ National Centre for Nuclear Research, Lodz, Poland}
\author{G Pr\'ev\^ot} 
\affiliation{ APC, Univ Paris Diderot, CNRS/IN2P3, CEA/Irfu, Obs de Paris, Sorbonne Paris Cit\'e, France}
\author{E Reali}
\affiliation{ INFN, Sezione di Roma Tor Vergata, Roma, Italy}
\affiliation{ Universit\`a degli Studi di Roma Tor Vergata, Dipartimento di Fisica, Roma, Italy}
\author[0000-0003-1624-1709]{G Romoli}
\affiliation{ INFN, Sezione di Roma Tor Vergata, Roma, Italy}
\affiliation{ Universit\`a degli Studi di Roma Tor Vergata, Dipartimento di Fisica, Roma, Italy}
\author{M Ricci} 
\affiliation{ INFN-LNF, Frascati, Italy}
\author{N Sakaki} 
\affiliation{ RIKEN, Wako, Japan}
\author[0000-0001-9457-1338]{K Shinozaki} 
\affiliation{ National Centre for Nuclear Research, Lodz, Poland}
\author[0000-0002-0233-9476]{J Szabelski} 
\affiliation{ National Centre for Nuclear Research, Lodz, Poland}
\author{Y Takizawa} 
\affiliation{ RIKEN, Wako, Japan}
\author[0000-0002-2224-1281]{G Valentini}
\affiliation{ ASI, Italian Space Agency, Rome. Italy}
\author{M Vrabel}
\affiliation{ National Centre for Nuclear Research, Lodz, Poland}
\author{L  Wiencke}
\affiliation{ Department of Physics, Colorado School of Mines, Golden, USA}

\correspondingauthor{M. Casolino, L. Marcelli}
\email{marco.casolino@roma2.infn.it, laura.marcelli@roma2.infn.it}

\begin{abstract}
Mini-EUSO (Multiwavelength Imaging New Instrument for the Extreme Universe Space Observatory) is a telescope observing the Earth   from the International Space Station since 2019. The instrument  employs a  Fresnel-lens optical system and a focal surface composed of 36 Multi-Anode Photomultiplier tubes, 64 channels each, for a total of 2304 channels with single photon counting sensitivity. Mini-EUSO also contains two ancillary cameras to complement  measurements in the near infrared and visible ranges \citep{TURRIZIANI20191188}. 
The scientific objectives of the mission range from the search for Extensive Air Showers (EAS) generated by Ultra-High Energy Cosmic Rays (UHECRs) with energies above 10$^{21}$~eV, the search for nuclearites and Strange Quark Matter (SQM), up to the study of atmospheric phenomena such as Transient Luminous Events (TLEs), meteors and meteoroids.       
Mini-EUSO can map the night-time Earth in the near UV range (predominantly between 290 - 430~nm) with a spatial resolution of about 6.3~km (full field of view equal to 44$^{\circ}$) and a maximum temporal resolution of 2.5~$\mu$s, observing our planet  through a  nadir-facing UV-transparent window in the Russian Zvezda module.
The detector saves triggered  transient phenomena with a sampling rate of  2.5 $\mu$s and  320~$\mu$s, as well as continuous acquisition at 40.96~ms scale. 
In this paper we discuss the detector response and the flat-fielding and calibration  procedures. Using the 40.96~ms data, we present $\simeq~6.3$~km resolution night-time Earth maps in the UV band, and report on various emissions of anthropogenic and natural origin.  
We measure ionospheric airglow emissions of dark moonless nights over the sea and ground, studying the  effect of clouds,  moonlight, and artificial (towns, boats) lights. In addition to paving the way forward for the study of long-term variations of light of natural and artificial origin, we also estimate the observation live-time of future UHECR detectors.  
 \end{abstract}


\section{Introduction} \label{sec:introduction}
Mini-EUSO (Multiwavelength Imaging New Instrument for the Extreme Universe Space Observatory, known as   {\textit{UV atmosphere} } in the Russian Space Program) is a  telescope operating in the  near UV range, predominantly between 290 - 430~nm, with a square focal surface corresponding to a square field of view of   44$^{\circ}$. Its spatial resolution at ground level is  $\simeq 6.3 \times 6.3\:$~km$^2$, varying slightly with the altitude of the International Space Station (ISS) and the pointing direction of each pixel. Mini-EUSO was launched with  the uncrewed Soyuz~MS-14, on 2019-08-22. The first observations, from the  nadir-facing UV-transparent window in the Russian Zvezda module, took place on 2019-10-07. 

The Mini-EUSO detector was originally designed for the development of the  study of  {Ultra} High Energy Cosmic Rays  { (UHECRs)} from space \citep{ptep}, as part of an ongoing effort by the JEM-EUSO (Joint Experiment Missions for Extreme Universe Space Observatory) collaboration.  So far, various instruments have been constructed and operated on the ground (EUSO-TA \citep{2018APh...102...98A}), on stratospheric balloons  (EUSO-Balloon \citep{EUSO-BALLLOON-Adams2022,2019APh...111...54A}, EUSO-SPB1 \citep{2017ICRC...35.1097W}  and EUSO-SPB2 \citep{ICRC__Eser:2021H6}) 
and in space (TUS \citep{Klimov2017}, in addition to the planned K-EUSO \citep{k-euso-universe} and POEMMA \citep{2019BAAS...51g..99O, Olinto_2021} missions).

The detector employs a BG3 filter (280 - 430~nm at 50\% transmission) on its focal surface to perform observations where most of the fluorescence light from Extensive Air Showers (EAS) initiated by cosmic rays interacting in the atmosphere is emitted (300 - 430~nm range) \citep{privitera-AVE200741}\footnote{The maximum $X_{\rm max}$ of a $10^{19}$~eV$\leq E\leq 10^{20}$~eV cosmic ray shower occurs at $750$~g/cm$^2$ $\leq X_{\rm max}\leq 800$~g/cm$^2$, corresponding to an altitude  between 2 and 2.5~km for a vertically incident event \citep{TAAbbasi_2018, Grieder2001CosmicRA}.}, with a $48 \times 48$~pixel {focal surface} ($\simeq~6.3$~km spatial resolution on the ground) and a sampling time of $2.5~\mu$s.  
 
In the case of EAS, the light is emitted by the return to the ground state  of nitrogen molecules (N$_2$) that are excited by the ionization of the charged particle component of the shower. 

Observations in this wavelength range with this combination of temporal and spatial resolution are relatively scarce and systematic observations of this type from space can contribute to study various phenomena that take place on the surface of the  planet or in its atmosphere, either with a terrestrial (e.g.~Transient Luminous Events, airglow, gravity waves)    \citep{2015ExA....40..239A, 2022icrc.confE.367M, gravitywaves}   or extra-terrestrial (e.g.~meteors, hypothetical strange quark matter) \citep{Adams2015} origin. Night sky emissions at wavelengths from 250 to 500~nm are mostly due to airglow generated from O$_2$:  the Oxygen molecules are separated into individual atoms by daytime solar radiation and then recombine to an excited state during night-time. Their subsequent deexcitation is the major source of emissions of Oxygen lines  
\citep{Airglowbook,MEIER_UV_1991SSRv...58....1M}. A recent measurement from the ground    estimates  the intensity of the airglow in the range 300 - 430~nm to be $1016\pm100~$ph$/($m$^2$~sr~ns$)$ \citep{MACKOVJAK2019150}. 

Outside the field of cosmic rays, where the night-sky emissions constitute a background to a transient EAS signal lasting $\simeq~100~\mu$s, night-time observations of the Earth in the UV band are important to a number of terrestrial, atmospheric, and space weather phenomena (see \citep{MEIER_UV_1991SSRv...58....1M} for an early review). Historically, most of the observations from space have involved the higher energy band of the spectrum ($<160$~nm) due to the information it provides on the structure and evolution of the ionosphere; for instance the data acquired by the OGO4 satellite \citep{OGO4} (UV nightglow at 130.4~nm and 135.6~nm), during the Apollo missions \citep{apollo16} (in the 105 - 160~nm and 125 - 160~nm ranges), {by} the NASA TIMED GUVI spectrographic imagers \citep{SIS},   {and the} TIMED/
GUVI \citep{2013TIMED} (135.6~nm). In the 300-400~nm   {range} we   { note} the observations performed by the UV detectors on board the Tatiana~\citep{tatianaGARIPOV2005400}, Universitetsky-Tatiana-2~\citep{Garipov2013} and Vernov~\citep{VernovSatelliteDataofTransientAtmosphericEvents} satellites. {In these experiments a quasi-stationary airglow, aurora emission, and anthropogenic sources were measured, as well as TLEs. More detailed studies of the space-time structure of various UV atmospheric emissions were conducted during the TUS experiment on board the Lomonosov satellite~\citep{rs11202449}.}

Given the focus of past efforts, the fast ($2.5~\mu$s) acquisition speed of Mini-EUSO and its high sensitivity can provide a unique data set for the study of atmospheric phenomena and terrestrial UV emissions, both natural (bioluminescence \citep{Miller14181}, ELVES, meteors, search for Strange Quark Matter), and anthropogenic (towns, fishing boats,  flashers). The inter-disciplinary possibilities of this type of detector have long been understood by the JEM-EUSO collaboration and so the instrument and mission have been designed with Earth observation capabilities in mind, {including} saving the data at a lower sampling rate (40.96~ms). 

In this paper  we report about Mini-EUSO observations of night-time near-UV terrestrial emissions, of natural or anthropogenic nature, focusing on emissions taking place on a timescale of 40.96~ms. The methods of data processing, flat-fielding and analysis are  also reported. With calibrated data we measure ionospheric airglow emissions of dark moonless nights over the sea and ground and the role of clouds and moonlight on the UV emissions.
We also report on the observations of anthropogenic light sources (towns, fishing boats) on the night-time UV environment.
We also evaluate  - in light of the aforementioned UV emissions - the observation live-time of Extended Air Showers generated by UHECR for future space-borne detectors.

\section{Instrument Overview}\label{InstView}
\begin{figure}[ht]
\centering
\includegraphics[width=0.8\textwidth]{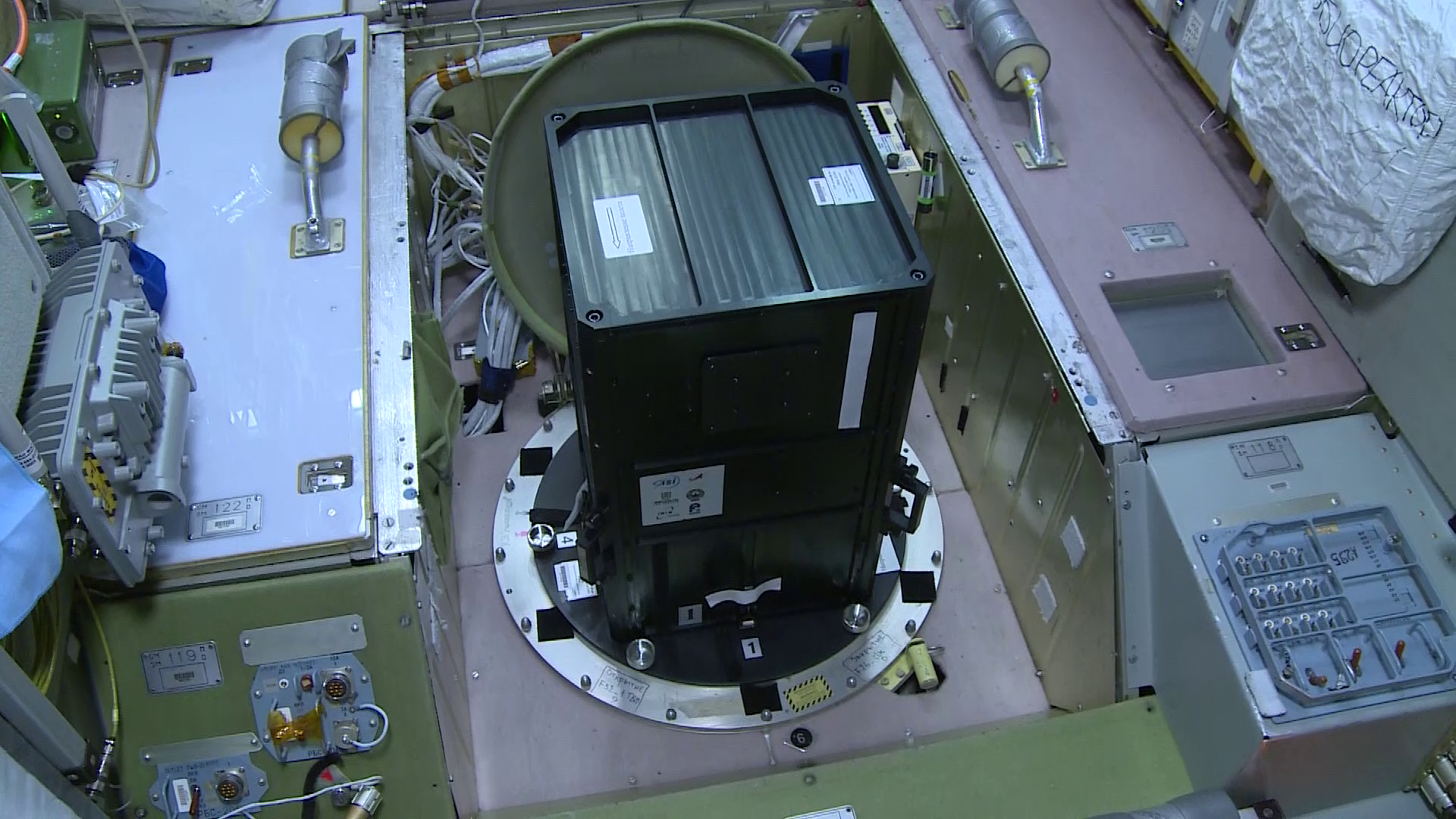}
\caption{Mini-EUSO installed inside the ISS on the UV-transparent window of the Zvezda module. The round porthole on the bottom of the picture looks nadir; the bottom center part (marked ``1'') is oriented toward  the direction of the velocity vector  of the station.}
\label{instrument_mounted}        
\end{figure}

Mini-EUSO \citep{Bacholle_2021} has been designed to operate from the interior of the ISS on the UV-transparent window located in the Zvezda module. The detector size ($37\times37\times62$~cm$^3$) was thus constrained by the dimension of the window and the Soyuz spacecraft. As all instruments operating in the ISS, the design is consistent with the safety  requirements (no sharp edges, low surface temperature, robustness...) associated with manned spaceflight.   Installation at the window is done  via a mechanical adapter flange; the only connection to the ISS is via a 28~V power supply and a grounding cable. The  weight of the instrument is about 35~kg, including the 5~kg flange, and the total instrument power consumption is $\simeq60$~W. Being located in the middle of the Zvezda module, the detector is usually installed during on board night-time, approximately at 18:30~UTC, with operations lasting about 12 hours until the following local morning (Figure~\ref{instrument_mounted}). To overcome the bottleneck of the limited telemetry flow from the station, data are handled  by the CPU \citep{Data_acquisition_Software} and stored  locally on 512~GB USB Solid State Disks (SSD) inserted in the side of the telescope by the cosmonauts at the start of each session.  Although no direct  telecommunication with the ground is present, after each  data-taking session, samples of data (about 10$\%$ of stored data, usually corresponding to the beginning and the end of each session)  are copied  and transmitted  to ground to verify the correct functioning of the instrument and subsequently optimize its working parameters. If necessary,  before each session, specific working parameters and patches in software and firmware are uplinked to the ISS and then copied on the SSD disk to fine-tune the acquisition of the telescope. Pouches, containing 25 SSDs, are then returned to Earth every $\simeq$~12~months  by  Soyuz spacecraft.

The optics consist of two 25~cm diameter Fresnel lenses in Poly(methyl methacrylate) -- PMMA. This material allows  for  a  light (11~mm thickness, 0.87~kg/lens), robust, and compact design well suited for space applications. 

\begin{figure}[ht]
\centering
 \includegraphics[width=0.6\textwidth]{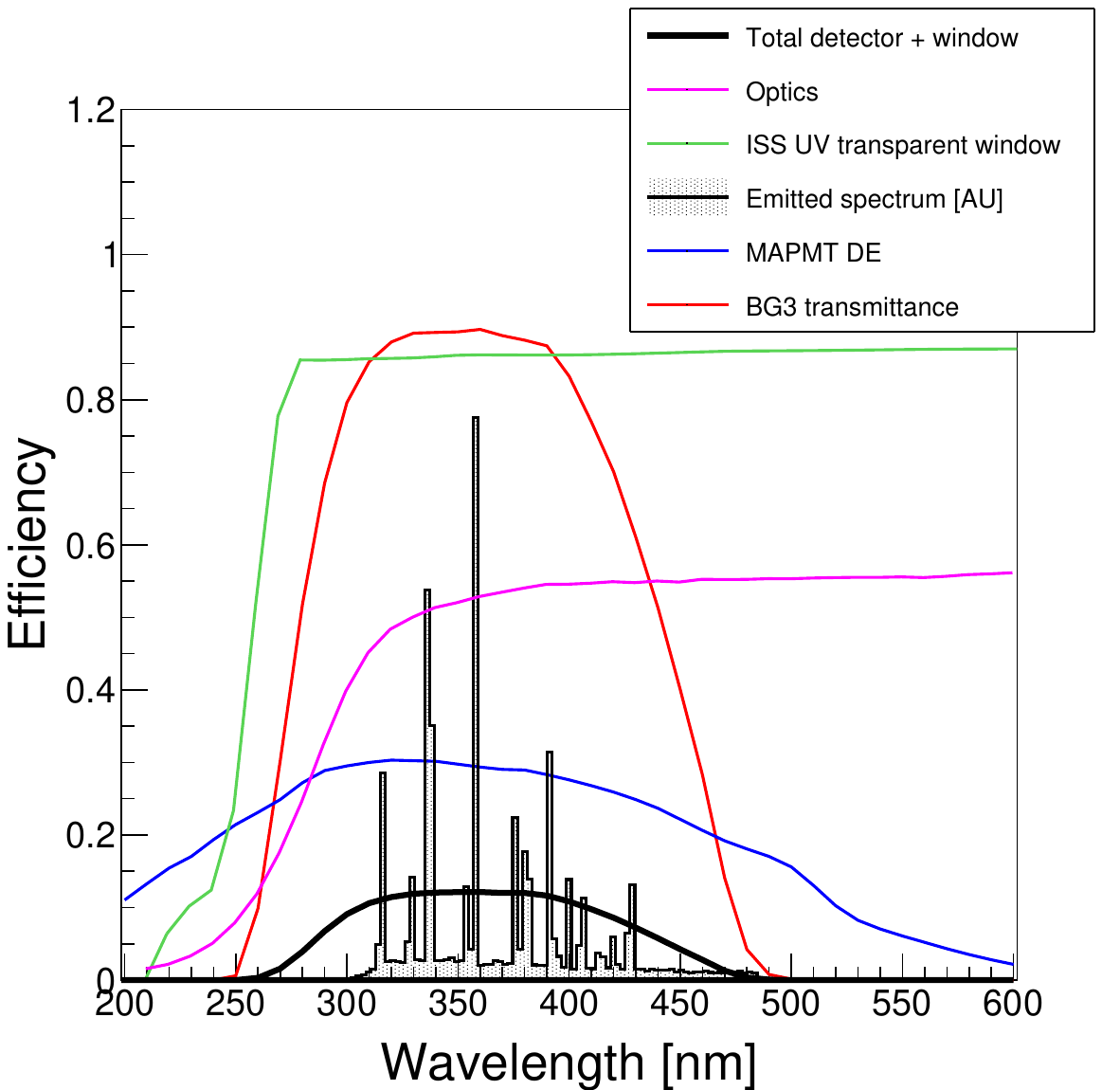}
\caption{The overall detection efficiency of the Mini-EUSO detector (black curve) as a function of wavelength. This is the  result of the  transmittance of the UV transparent window of the ISS (green curve), the optics  (purple curve), the BG3 bandpass filter (red curve) and the MAPMT photon detection efficiency (blue curve). The system has been designed to maximize observations of the fluorescence light emitted by nitrogen molecules excited by the extensive air shower of cosmic rays (grey histogram).
Mini-EUSO has a total efficiency higher than 50\% of the maximum in the wavelength range 290-430~nm.}
\label{efficiency}         
\end{figure}

The Mini-EUSO Focal Surface (FS), or Photon Detector Module (PDM), consists of a matrix of 36 Multi-Anode Photomultiplier Tubes (MAPMTs, Hamamatsu Photonics  R11265-M64), arranged in an array of 6$\times$6 elements. Each MAPMT consists of 8$\times$8 pixels, resulting in a total of 2304 channels.   The MAPMTs are grouped in Elementary Cells (ECs) of $2\times2$ MAPMTs. Each EC has an independent high voltage power supply and board connecting the dynodes and anodes of the four photomultipliers.
Each EC (250~g per EC, including filters, MAPMTs and the High Voltage Power Supply - HVPS) is  potted with Arathane to avoid electrostatic discharges and short circuits from microscopic floating debris. Each MAPMT in the EC is read by an ASIC that performs single photon counting and sends the data to the FPGA (Zynq) board.   
The effective focal length of the system is 205~mm (focal length 300~mm), with a Point Spread Function (PSF) of 1.2 MAPMT pixels. 
UV bandpass filters (2~mm of BG3 material) with anti-reflective coating are glued in front of the MAPMTs  to predominantly select wavelengths between 290~nm and 430~nm.  In Figure~\ref{efficiency} are shown the various contributions to the overall detector efficiency. The Detection Efficiency (DE) of the MAPMTs has been obtained by rescaling the Quantum Efficiency (QE) curve provided by Hamamatsu by a typical collection efficiency of 80$\%$. The result is in good agreement with the detection efficiency measured in the laboratory at 398~nm.
 
Similar PDM units have been employed in the ground-based telescope of EUSO-TA \citep{2018APh...102...98A} and  in the first two balloon flights: EUSO-Balloon \citep{2015ExA....40..281A,2019APh...111...54A,EUSO-BALLLOON-Adams2022}  and EUSO-SPB1 \citep{2017ICRC...35..384B}. A larger focal surface with three PDMs side-by-side will be used in the upcoming EUSO-SPB2 flight \citep{ICRC__Eser:2021H6}. 

Prior to the launch, the instrument underwent a series of integration and acceptance tests \citep{refId0, Bisconti_preflight} in Rome, Moscow, and Baikonur cosmodrome, where it  was  integrated in the uncrewed Soyuz capsule. The launch took place on 2019-08-22 and the telescope was first turned on on 2019-07-10 (Figure~\ref{instrument_mounted}) when the first trained crew reached the station.   
The commissioning phase required a gradual operation of the instrument:  the  first session involved operation in safe mode, with only one EC unit active and with a lower voltage (corresponding to a photon sensitivity of about 1$\%$). Over the course of the following sessions, the  acquisitions used the full PDM in lower voltage mode. Subsequently, the full focal surface was turned on at the nominal voltage, and the software and firmware parameters have been fine-tuned to optimize the acquisition. See \citep{Mini-EUSO_trigger, Data_acquisition_Software} for a description of the software and the acquisition procedures and \citep{Bacholle_2021} for  details on the detector and the data gathered so far.

\section{Data Handling}\label{sec:methods}
Mini-EUSO features a multi-level data acquisition system \citep{Mini-EUSO_trigger, Data_acquisition_Software} capable of simultaneously storing events at three different timescales. This allows the observation of the same event with three different time resolutions:
\begin{itemize}
\item {\bf 2.5~$\mu$s} time resolution: this timescale features a dedicated trigger system. A single frame is called D1 GTU (Gate Time Unit). This is the acquisition speed of the system: every D1 GTU data are read by the 36 ASICs (one per MAPMT) and sent to the PDM FPGA acquisition board for  acquisition and processing. Data are stored by the FPGA in a circular buffer of 128 GTUs. When the trigger conditions are met, the data of the preceding  64 GTUs and the following 64 GTUs are stored and then sent to the CPU.  \textcolor{black}{ The trigger algorithm, described in \citep{BATTISTI20222750}, looks for fluctuations $16 \sigma$ above the average value (dynamically updated) in each pixel. The excess signal must persist for more than 8 GTU (20 $\mu s$) in any given pixel. This is the time needed for  a  EAS-generated signal to cross the field of view of each pixel (6.3~km) at the speed of light. The $16 \sigma$ threshold is required to keep low the rate of random occurrences due to non-Poissonian fluctuations from noisy pixels (very noisy pixels can also be masked by the FPGA logic). As will be discussed in section \ref{sec:UHECR}, the lens size limits the energy threshold to   particles above $\simeq~10^{21.5}$~eV, that so far have not been observed with ground-based observatories (nor with our detector). The algorithm was found to work correctly, triggering on ground Xenon flashers, that exhibit a signal similar to that from UHECRs, but with a light curve that allows them to be identified as artificial lights. This algorithm has also been successful in  triggering on Transient Luminous Events such as Sprites and ELVES \citep{Bacholle_2021}.  }

\item {\bf 320~$\mu$s } time resolution: this timescale features a dedicated trigger system. A single frame is called D2 GTU. Each of these frames is the sum of 128 D1 GTUs and is calculated by the PDM acquisition board. If the trigger conditions are met, data are stored in a similar manner to D1 GTU and sent to the CPU.

\item {\bf 40.96~ms }time resolution: data frames are saved continuously without a trigger system and sent to the CPU. A single frame is called D3 GTU. Each of these frames is the sum of $128\times 128 = 16384$ D1 GTUs) and is calculated by the PDM acquisition board.  This timescale is the only one without a trigger system and performs a continuous monitoring of the UV emission of the Earth. It is  used for the observation of meteors, the search for Strange Quark Matter, and for mapping of the night-time terrestrial UV  emissions.  
\end{itemize}  

In this work all values have been rescaled to give the corresponding number of counts per 2.5~$\mu$s (one D1 GTU), henceforward simply called a GTU. This means that a pixel which we report as showing 1 count/GTU actually detected 128$\times128 = 16384$~counts in 40.96~ms (one D3 GTU), averaging to 1 count every 2.5~$\mu$s.

\subsection{Dead Time and large signal non-linearity}
The detector readout is performed by Spaciroc-3 ASICs \citep{Blin:2018tjp}. Each ASIC has 64 channels and counts the number of single photoelectrons signals arriving in one $2.5~\mu$s GTU window\footnote{Each D1 GTU has a dead time of 50~ns, needed for the ASIC to process the data. The active time is thus $2.45~\mu$s. This dead time has been taken into account in all data processing and calculations, but for simplicity we refer to 2.5 $\mu$s throughout this paper.}. The response of the ASIC to a single photoelectron results in a dead time of $\tau_0= 5~$ns, and therefore if two or more photons arrive within this interval, only one is counted. Assuming that the  distribution of the time of arrival of the photons follows Poissonian statistics, the relationship between  the estimated number of photoelectrons~(pe) hitting the pixel, $n_\textrm{pe}$, per given GTU and  the detected photoelectrons,  {$n$}, is ~\citep{Pomme08}:
 
\begin{equation}
  n = n_\textrm{pe} e^{\left(-n_\textrm{pe}\cdot\frac{\tau_0}{\tau_\textrm{GTU}}\right)}
\end{equation}

\begin{figure}[h]\centering
\centering
\includegraphics[width=0.95\textwidth]{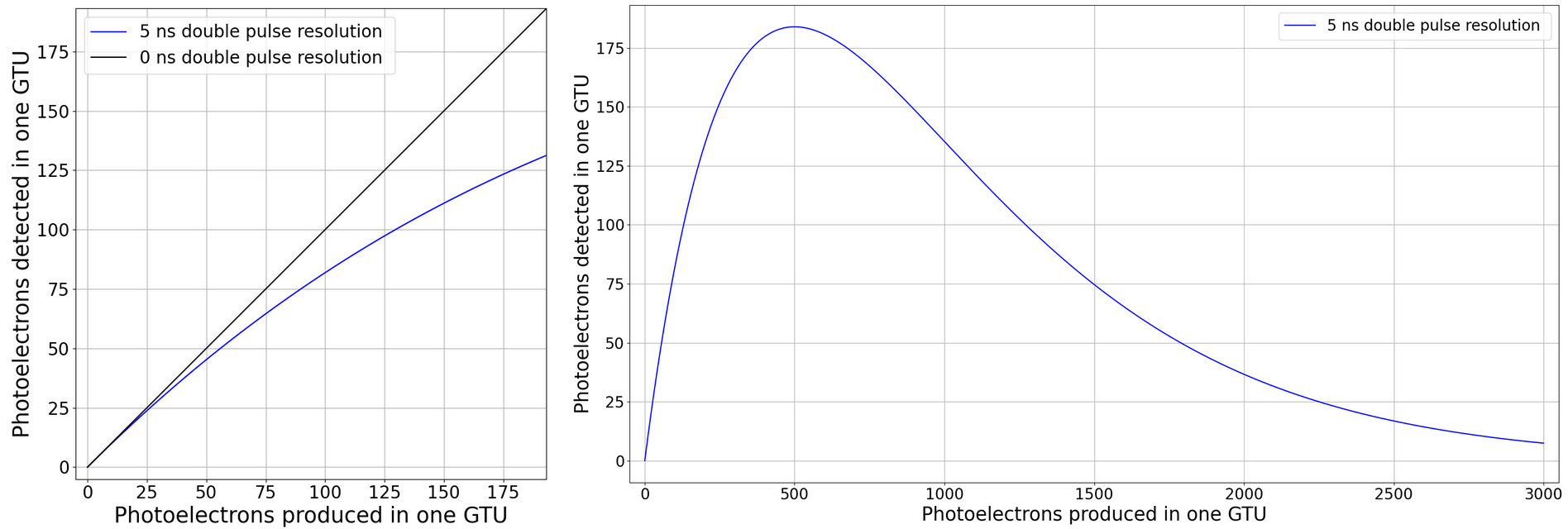} 
\caption{Detected photoelectrons in one 2.5~$\mu$s GTU, $n$, as a function of produced photoelectrons, $n_\textrm{pe}$. The difference is due to the pile-up assuming a dead time in the ASIC of 5~ns after each detected photoelectron. }
\label{pileup}
\end{figure}

The difference between $n$ and  $n_\textrm{pe}$ is negligible for small counts  but grows with intense signals (see Figure~\ref{pileup}).  The maximum   { $n$} is $\sim 180 $/GTU which corresponds to { $n_\textrm{pe,PMT}$} $\sim $500/GTU photoelectrons produced by the MAPMT cathode. For higher values, the detected  signal decreases and thus the true value of $n_\textrm{pe}$ can only be determined if the order of magnitude of the signal is known (e.g.~from nearby pixels). Anyway, in most cases, the signals are much lower than this amount and reach saturation only in the case of very bright lightning which is usually discarded during the analysis\footnote{Usually these events have a spatial halo of lower intensity that allows us to remove the saturation effect.}.
This correction is applied {offline} directly to D1 triggered values, since we read the number of counts/GTU. However, for D2 (320$~\mu$s) and D3 (40.96~ms) data, the averages performed in real time during the acquisition do not allow us to take this correction into account, resulting in a lower value being detected. A correction can still be applied assuming that the average light detected is uniformly spread in the D2 or D3 GTUs which compose each frame; however this is usually not needed, since bright lights also trigger safety mechanisms that alter the detector response.

\subsection{Safety mechanisms for MAPMT protection}

\begin{figure}[h]\centering
\centering
\includegraphics[width=0.9\textwidth]{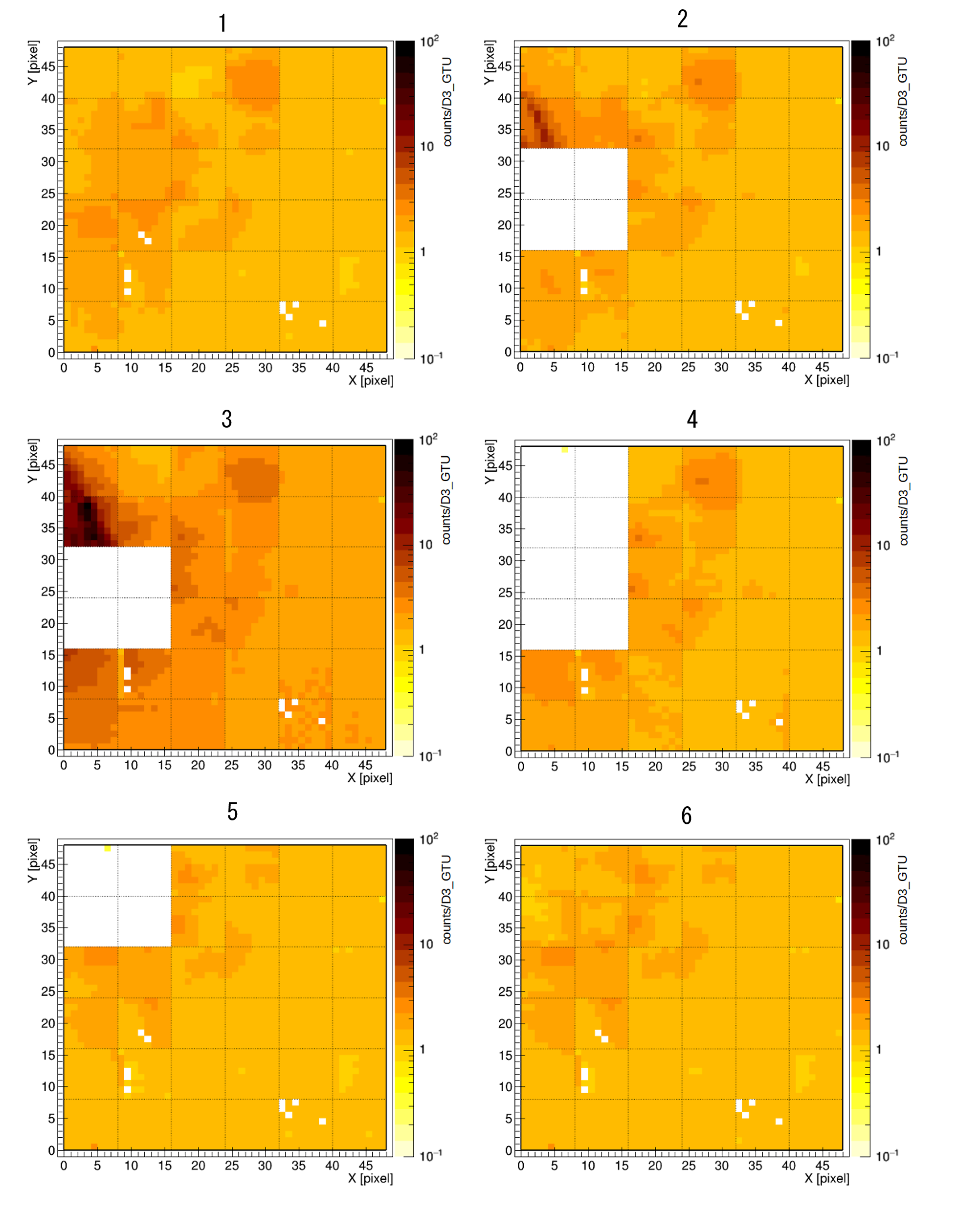}
\caption{From Top left to Bottom Right: During acquisition (2020/03/03, 00:51:22 GMT), a bright signal above 100~counts over more than 3 pixels in a GTU   triggers the safety mechanism of the HVPS (2020/03/03, 00:51:23 GMT). The corresponding EC-unit of 4 MAPMTs has the voltage difference of the photocatode and the first dynode  reduced to zero. In the picture, two ECs (first the centre-left, Panel 2-3 and then the top left, Panel 4-5) are thus set to zero by  the lightning-induced bright light.  After 244 D3 GTUs (about 10~seconds) the voltage is fully restored, since the brightness has decreased (panel 6, 00:51:33 GMT).}
\label{switches}
\end{figure}

The high voltage system has an internal analogue safety circuit which limits the maximum current drawn by an EC-unit to 1.6~$\mu$A. This protects the MAPMTs from being damaged by a too bright illumination  and is sensitive mostly to light spread-out over several pixels or MAPMTs. An additional  safety measure -  of digital nature - is implemented in the PDM board and is triggered by concentrated, very bright light sources. If more than 100~counts are present on more than three pixels of the same EC in a given GTU, the Zynq board activates the safety circuit on the  EC corresponding to the bright light, and the electric potential difference between the photocathode and the first dynode of the photomultiplier is removed. This   causes a drop in light collection efficiency by several orders of magnitude. After about 10 seconds in this mode (244 D3 GTUs), if the brightness is not above the previously mentioned threshold,  the full functionality is restored. The effect of this safety cycle is visible in Figure~\ref{switches}.

In most cases the signal is either not intense enough to require correction in D3 mode, or it is so bright that it triggers the safety mechanism. This usually happens during lightning events or very bright moonlight close to the zenith, but also very populated areas can emit enough light to trigger the safety mechanism within an EC.

\subsection{Flat Fielding} 

\begin{figure}[h]\centering
\centering
\includegraphics[width=.9\textwidth]{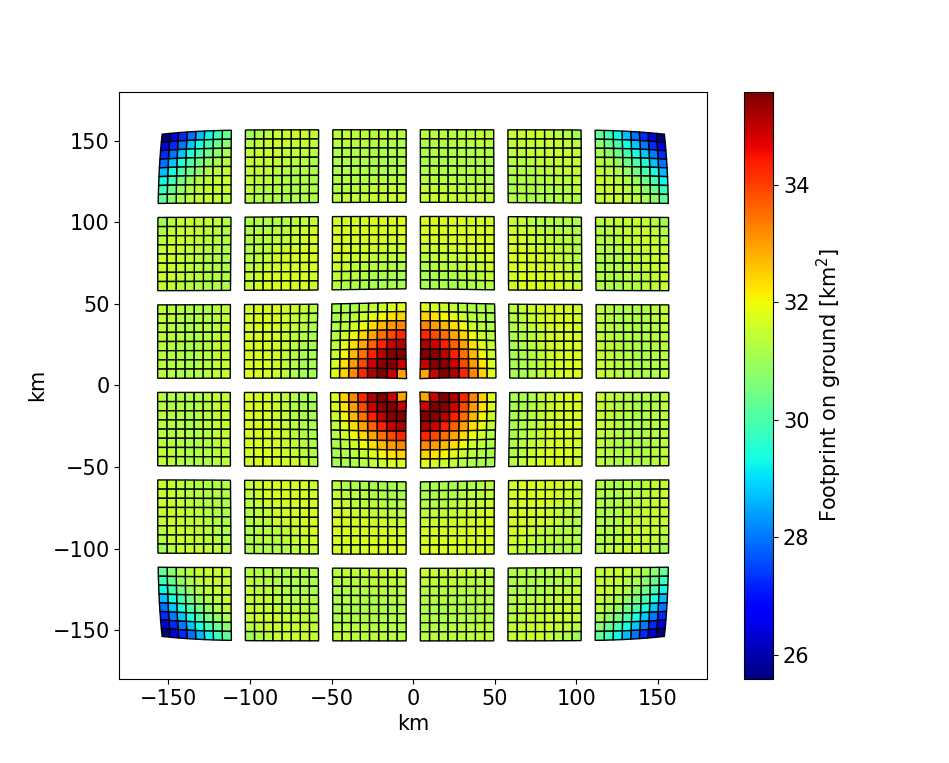} 
\caption{Footprint on the ground of each pixel of the focal surface, assuming the ISS is at an altitude of 400~km. The presence of gaps between MAPMTs and ECs is visible, along with the distortion produced by the optics at higher angles. The color scale represents the area on the ground observed by each pixel. These differences in area are taken into account in the flat fielding procedure. \textcolor{black}{Given the periodic variations in the height of the ISS, both due to eccentricity of the orbit, the long-term atmospheric dragging and its  consequent  reboost, the instantaneous field of view on the ground is continuously recalculated. The  curvature of the Earth has a negligible effect  on the field of view.} }
\label{fig:Pixels_dimension}
\end{figure}

Pixels within the detector which receive the same amount of light will not return the same count value due to variations in the MAPMT sensitivity (quantum efficiency, photoelectron collection efficiency, and gain) and in the readout electronics, both pixel-to-pixel within a single MAPMT and between different MAPMTs.
In addition, the optical system does not provide identical transmittance across the field of view nor a equal angular field of view for all pixels. In particular, pixels toward the edge of the field of view receive less light (i.e.~\textit{vignetting}) and show a modified footprint on ground (i.e.~\textit{barrel distortion}) compared to those in the center of the detector, as shown in Figure~\ref{fig:Pixels_dimension}. 
Furthermore, the BG3 filters mounted on each MAPMT are  cut in a truncated pyramid shape to reduce the amount of dead space present between adjacent MAPMTs\footnote{This is  done to increase the detected signal in case of UHECR (Ultra-High-Energy Cosmic Ray) atmospheric showers.}, and therefore pixels on the edges of a MAPMT collect light from a slightly larger focal plane area. The overall result of all these effects is that the light collected by each pixel depends on the position of the pixel inside the PDM.

\begin{figure}[h]\centering
\centering
\includegraphics[width=.9\textwidth]{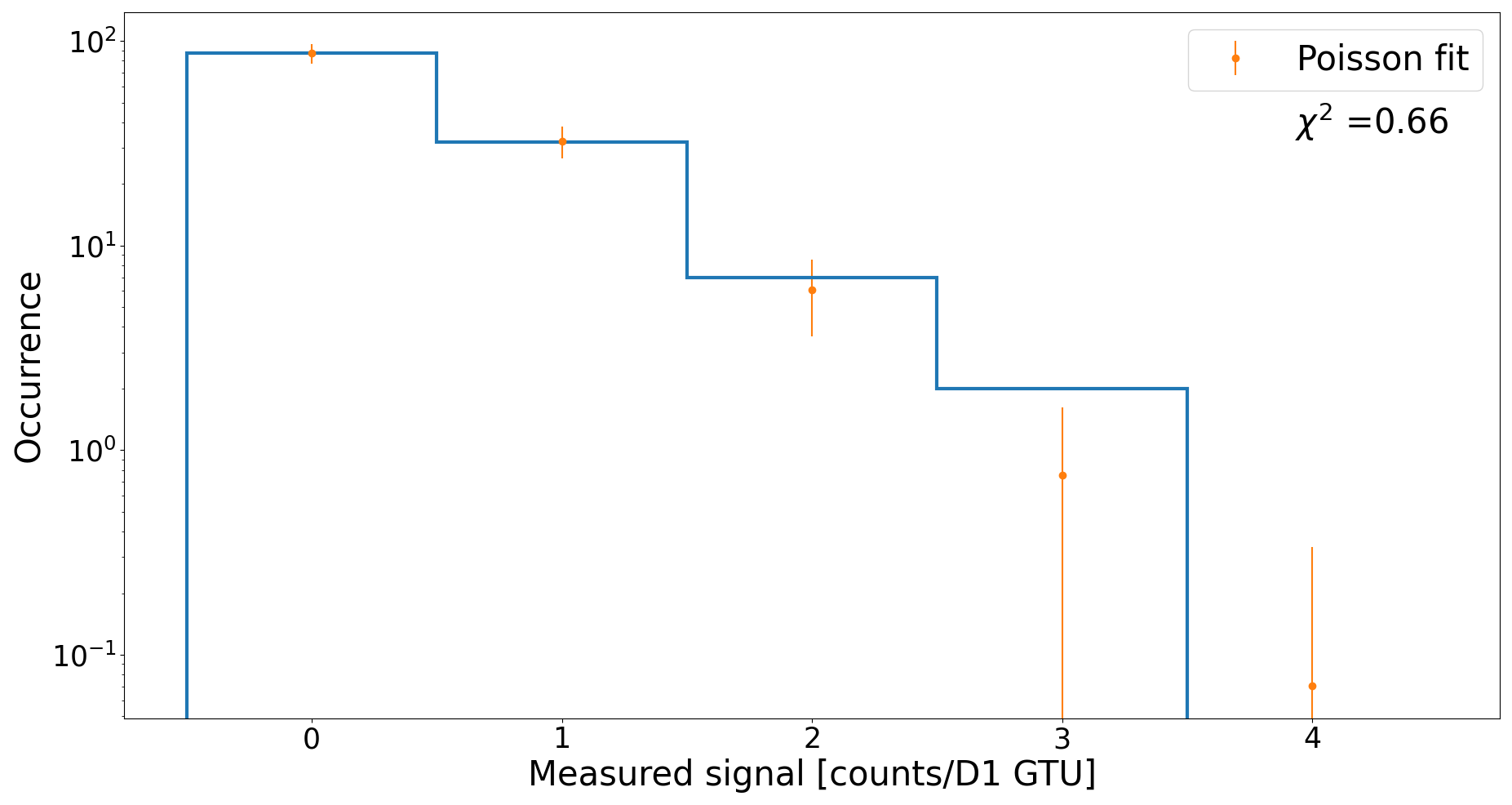} 
\caption{Histogram of the counts registered by a non-triggering pixel in 128 consecutive D1~GTUs and the Poissonian fit (orange points, mean value $\mu = 0.41$~counts/GTU).   
}
\label{poissonianstatsd1}
\end{figure}

The fluctuations between GTUs  of the photoelectron counts of each pixel are described by Poisson statistics: Figure \ref{poissonianstatsd1} shows an example of the distribution of the number of counts in one pixel observing a dark region on the ground, which agrees well with a Poisson distribution with a mean of $\mu = 0.41$~counts/GTU.

To compensate for the variation in pixel sensitivities within the PDM, we implement a ``bootstrapped" flat fielding procedure based upon three assumptions:
\begin{itemize}
    \item The conditions on the ground  which produce the lowest count values along an orbit are the same for all pixels (either ocean, desert, forest, ...).
    \item The difference in the number of counts produced by this environment is only due - in first approximation -  to the difference in the effective pixel sensitivities.
    \item All pixels observe this ``minimum light'' environment at some point in the session, even if only for a few GTUs.
\end{itemize}
The first two assumptions are generally true, while the third condition is justified by the fact that the ISS moves at $\sim$7.66~km/s, and therefore it takes slightly more than 20 D3 GTUs for the Mini-EUSO field of view to move 6.3~km and completely change a given pixel view. 
 
Moreover, the dimmest environment is typically represented by open ocean, deserts, or forests, all environments which span hundreds of km and are thus oberved at least once by all the pixels of the PDM.

\begin{figure}[h]\centering
\centering
\includegraphics[width=.99\textwidth]{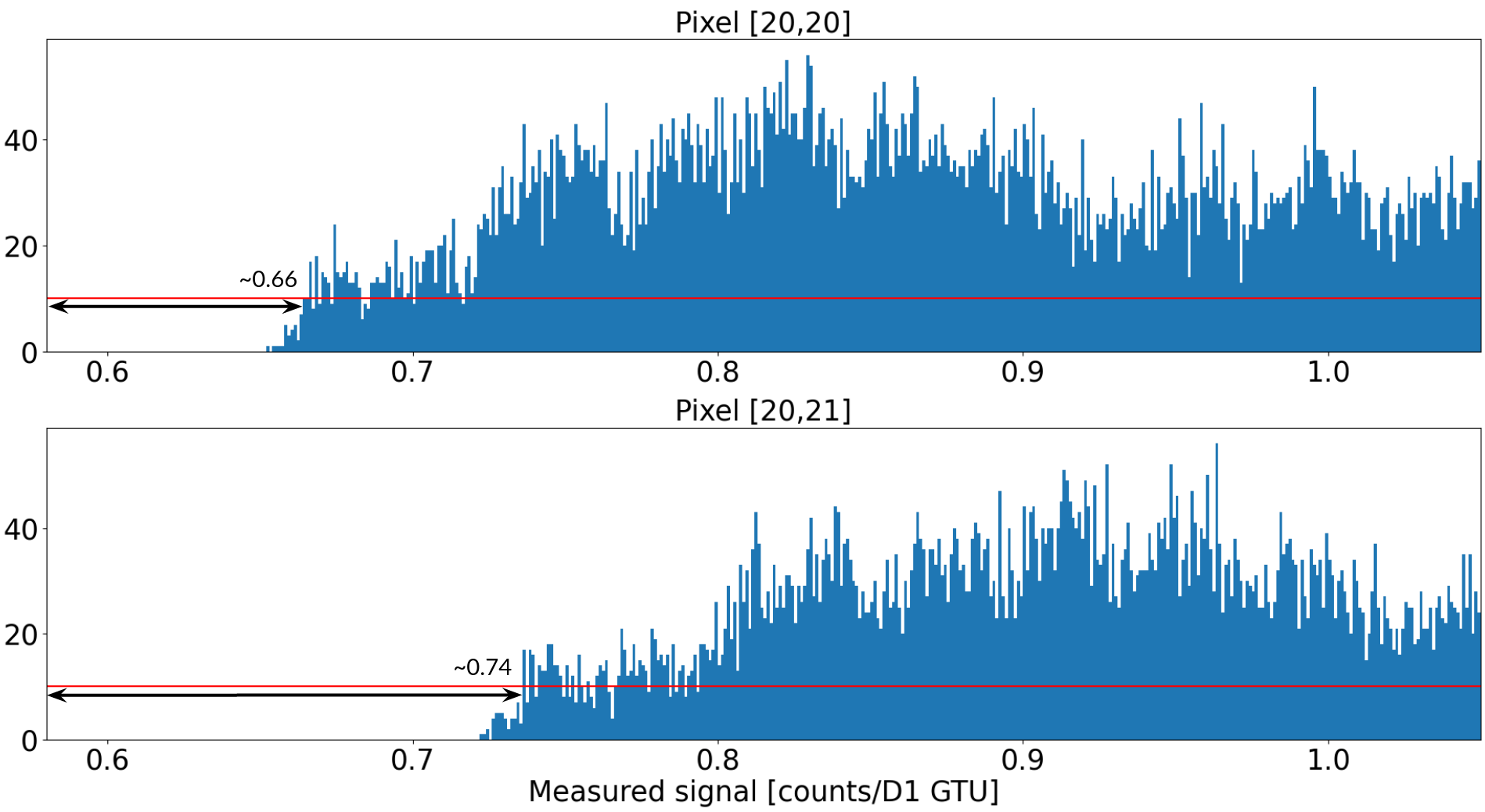} 
\caption{Histogram of the counts registered by two neighboring pixels (top panel: pixel (20,20); bottom panel: pixel (20,21)). Being next to each other, they observe almost the same environment. Nevertheless, pixel (20,21) (lower panel), presents higher counts than the other. The lowest bin populated with at least 10~counts (red horizontal line) is at $N_{\rm min}^{(20,20)}\simeq0.66$~counts/GTU for pixel (20,20) while it has an higher value of $N_{\rm min}^{(20,21)}\simeq0.74$~counts/GTU for pixel (20,21). The two different responses are normalized to unity by dividing the counts of each pixel in each GTU by these factors,  as can be seen in Figure~\ref{fig:Flat_field_V4_Lightcurves}.}
\label{fig:Flat_field_V4_Histograms}
\end{figure}

\begin{figure}[h]\centering
\centering
 
\includegraphics[width=.99\textwidth]{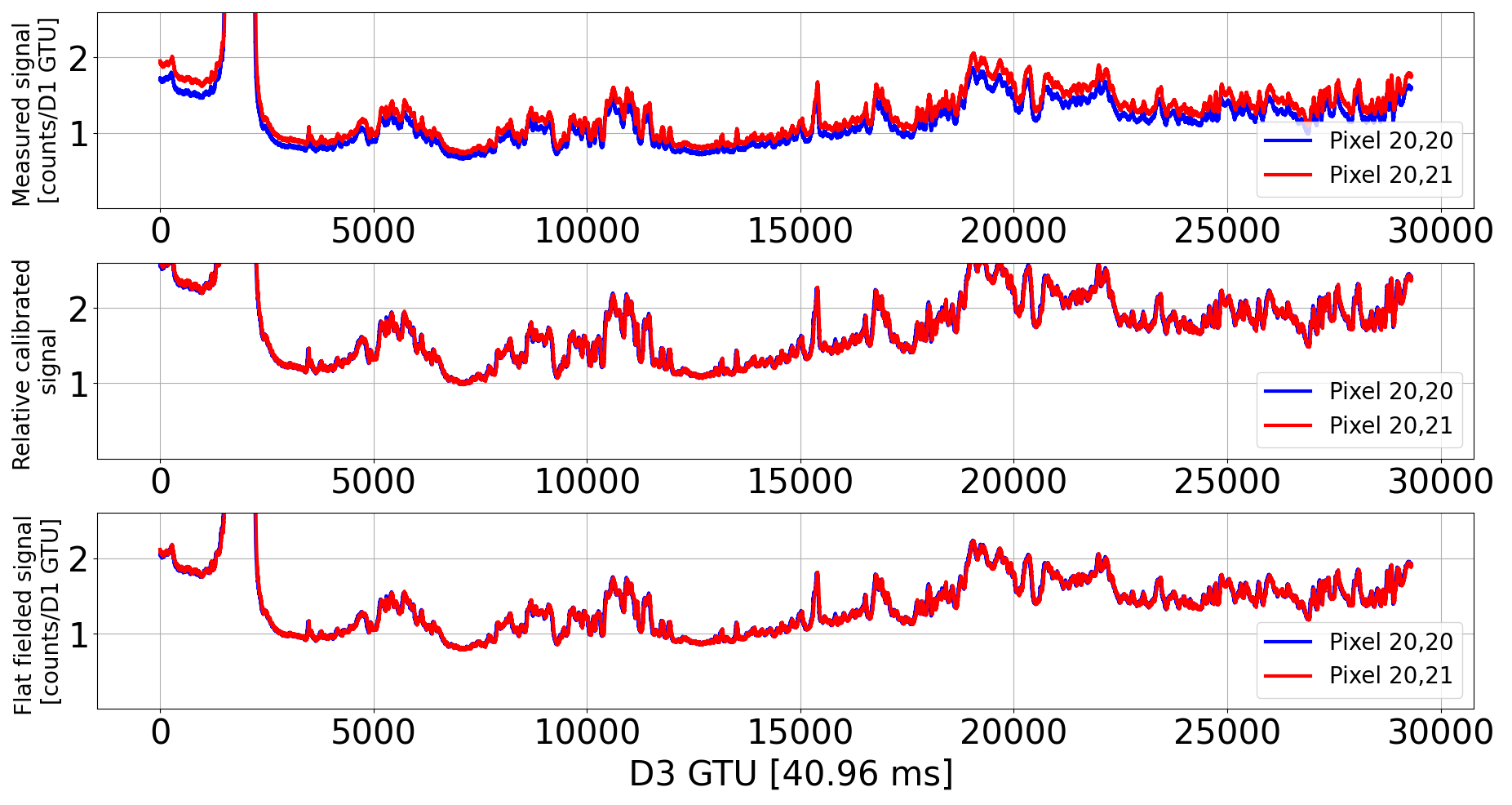} 
\caption{The lightcurve of the two neighboring pixels considered in Figure \ref{fig:Flat_field_V4_Histograms} as a function of time. \textbf{Top panel:} uncorrected signal data. It is possible to see that pixel (20,21) presents a response  higher than pixel (20,20) by the factor $N_{\rm min}^{(20,21)}/N_{\rm min}^{(20,20)}$; \textbf{Middle panel:} data after applying  the relative calibration (first step of the flat fielding procedure).  In this case, the two curves are more similar in behaviour but still require setting the session-dependent baseline.  \textbf{Bottom panel:} flat-fielded data, after baseline correction has been applied (second step) and the information on the average light level is restored.}
\label{fig:Flat_field_V4_Lightcurves}
\end{figure}

Under these assumptions, the minimum value which the ($i$th, $j$th) pixel registers over the session, $N_{\rm min}^{ij}$, can be used as a relative calibration factor for that pixel. We can then determine the flat field correction from the observed data itself in two-steps:
\begin{itemize}
    \item[1] Determination of the relative response difference between pixels, effectively normalizing the minimum response of each pixel to unity.
    \item[2] Calculation of a scale factor to recover the absolute count values for that  session. 
\end{itemize}
Reading the D3 data of an entire orbit (usually from 30 thousand to 60 thousand D3 GTUs), we identify and exclude any data not acquired at the nominal high voltage mode (e.g.~where the safety mechanisms have triggered to protect an EC from very bright illumination)\footnote{An EC is considered in lower voltage mode (\textit{cathode 2}) if there are more than 15 pixels with less than 0.001~counts in a single GTU as a result of the lower gain due to the removal of a dynode from the amplification chain.}. 
For each pixel, we then produce a histogram of the counts over the entire orbit with the bin width fixed to 0.001~counts, examples of which are shown for two neighboring pixels in Figure~\ref{fig:Flat_field_V4_Histograms}.
The lowest populated bin    of the  histogram of a given pixel, represents its    observation of the same minimum light environment, different due to the differences in pixel sensitivity. We define the lowest bin with at least 10~counts as the minimum signal $N_{\rm min}^{ij}$ (indicated by the black arrow in Figure~\ref{fig:Flat_field_V4_Histograms}). Division of the counts in each pixel by its respective value of $N_{\rm min}^{ij}$ provides relative calibration off all pixels across the PDM. 
 
After this relative calibration, the absolute scale of the flat-fielded counts no longer corresponds to the absolute amount of light observed during the orbit (see top and middle panels in Figure~\ref{fig:Flat_field_V4_Lightcurves}). For example, two orbits which observe the same uniform environment, such as the ocean, one on the night of a new Moon and one on a night with a Moon phase close to 50\% or higher will encounter very different values, since the background is highly dependent on the Moon phase. 
The relative flat fielding, however, scales the base-line of both orbit to 1, and therefore we must scale by a factor $K_{\rm abs}$ the count values according to the overall brightness of the orbit. 
Using a subset of 40 different relative calibrations,  we compute  the average and standard deviation of the $N_{\rm min}^{ij}$ values for all PDM pixels, and select a subset of  pixels which display a low standard deviation ($\frac{\text{Standard deviation}}{\text{Average}} \lesssim 5\% $)  as a representative stable sample from which to compute the absolute scale for each session.
The value of $K_{\rm abs}$ determined in this manner changes little when varying the number of pixels included in the sample, and the results obtained taking into account all 2304 pixels usually differs from that obtained considering only the {selected} sample by $\sim$ 10\% - 20\%.

\begin{figure}[h]\centering
\centering
\includegraphics[width=.37\textwidth]{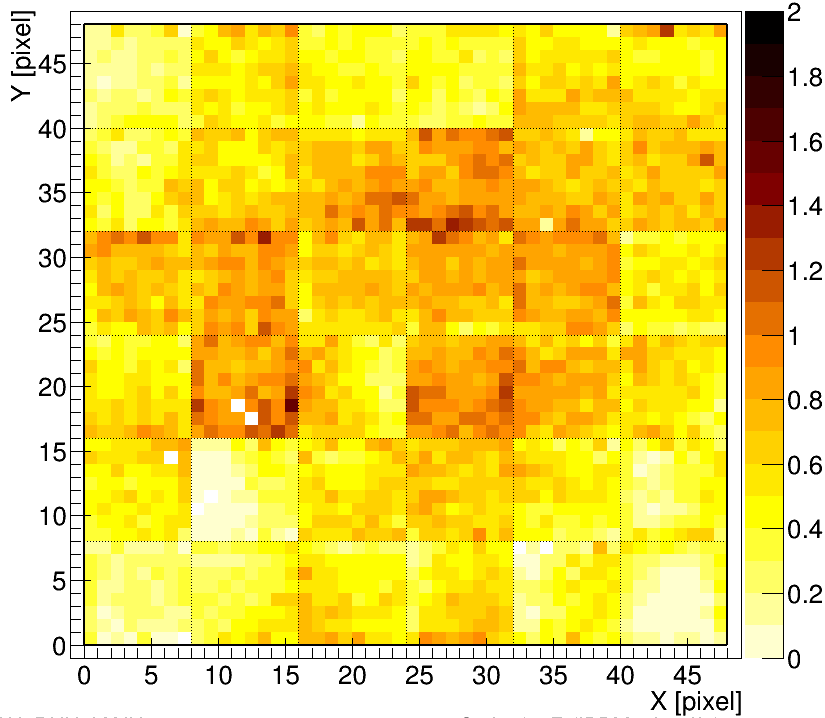} 
\includegraphics[width=.61\textwidth]{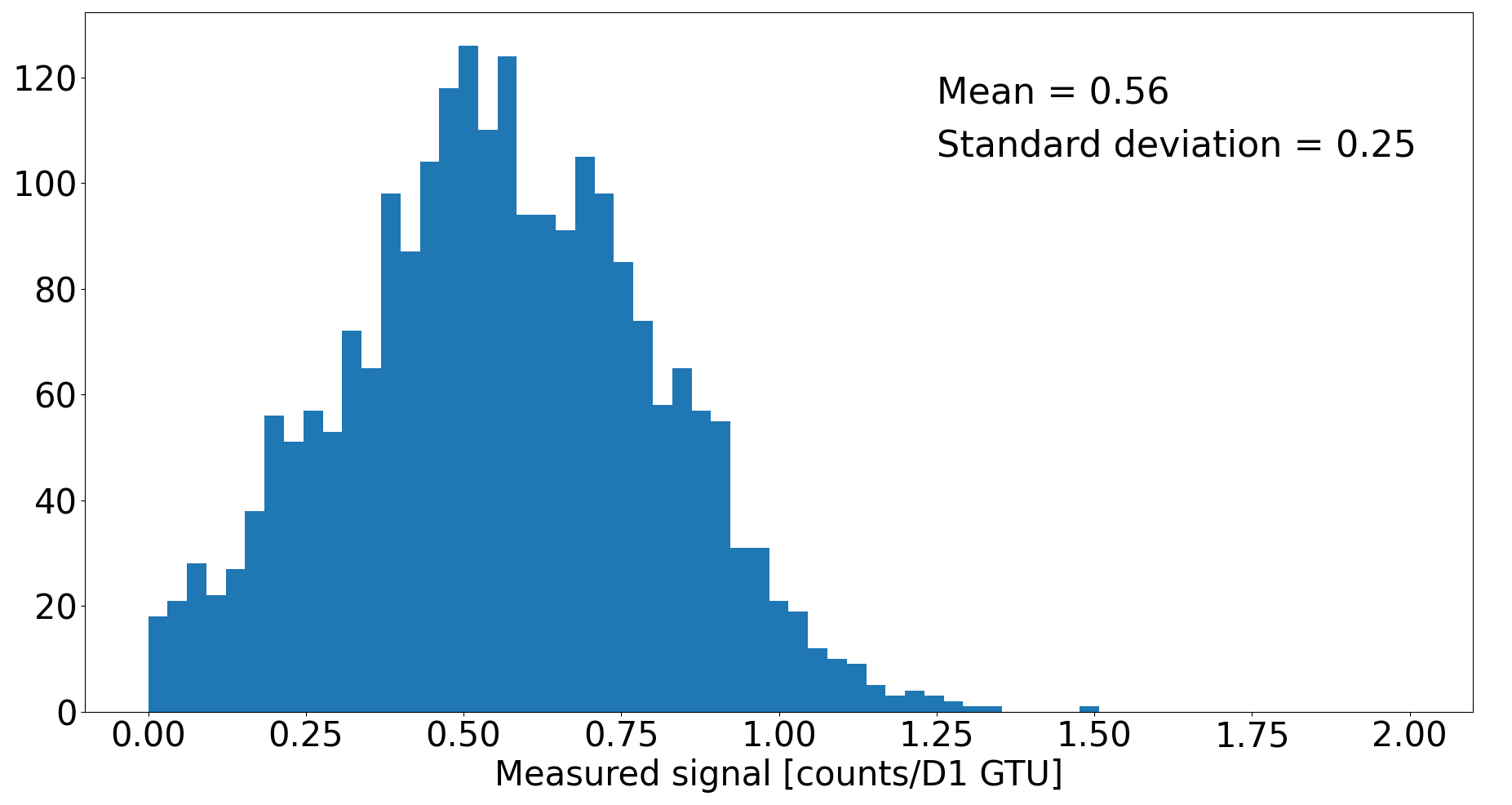} 
 
\includegraphics[width=.37\textwidth]{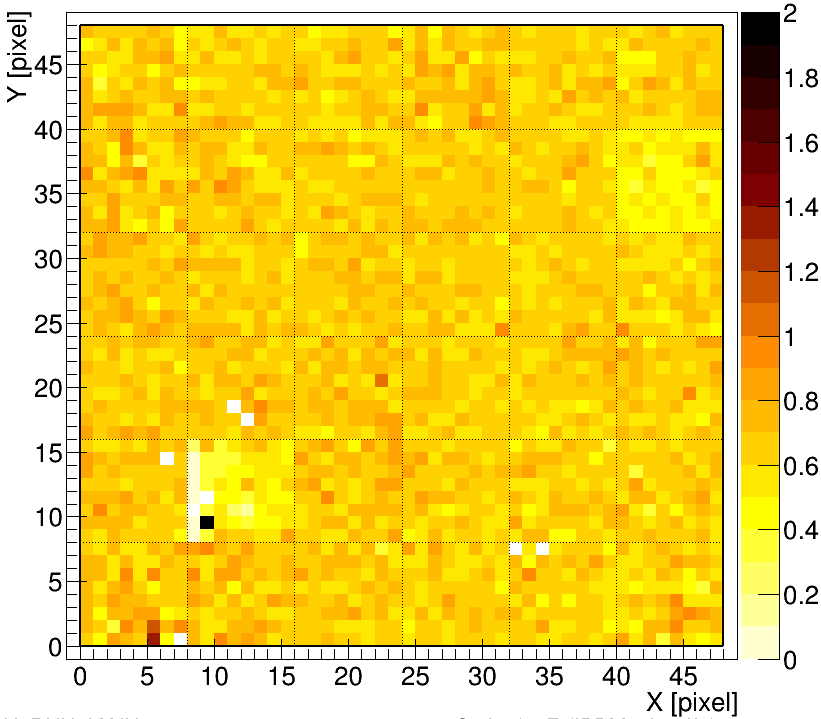} \includegraphics[width=.61\textwidth]{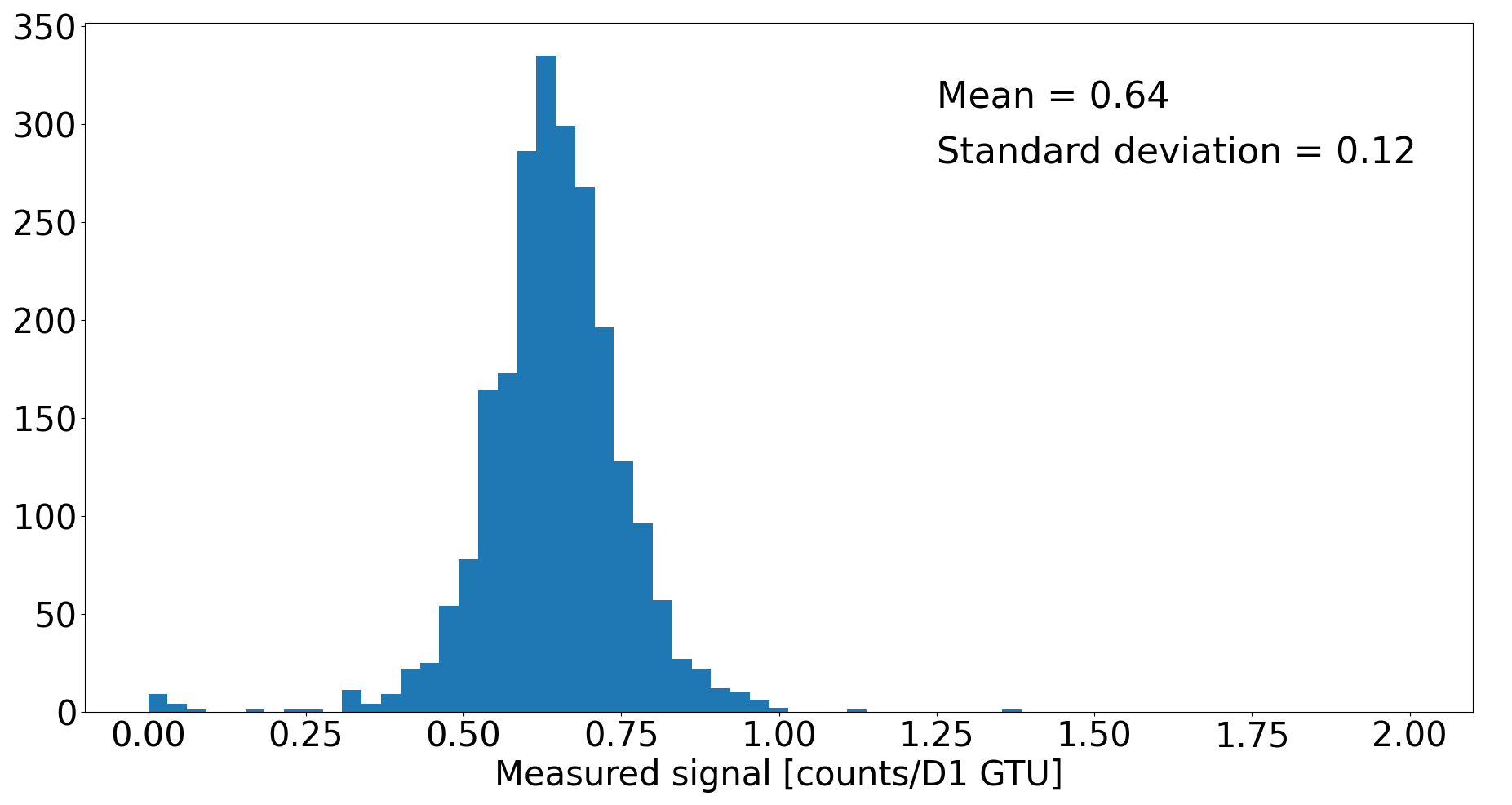} 
 
\caption{\textbf{Top Left:} average of 128 D1 GTUs (raw, unprocessed data) of Mini-EUSO observing the sea (Pacific Ocean region). Assuming a uniform emission due to airglow and the absence of additional light sources, the different counts in each pixel (histogram in top right panel) are mainly due to the different sensitivities of the pixels. 
\textbf{Bottom Left:} the same average  shown above, after the flat fielding procedure. The response of the pixels in the PDM is much more uniform. The two slightly dimmer MAPMTs (second row, sixth column, and fifth row, second column) are affected by a bit-shift error in the ASIC readout, which is corrected in preprocessing but still results in a lower efficiency, especially for low signals). The brightest pixel (X=9, Y=9) is the one that issued a trigger, in this case due to a low energy cosmic ray directly directly hitting the focal surface. There are also five non-functioning pixels. 
\textbf{Bottom Right:} histogram of the flat-fielded averages. The spread is significantly reduced, from \textcolor{black}{$0.25$ raw counts/GTU, to $0.12$ flat-fielded counts/GTU}.  
}
 
\label{fig:Flat_field_pacific_ocean}
\end{figure}

The final flat-fielded counts in each pixel are then given by
\begin{equation}
    N_{\rm flat-fielded}^{ij} = K_{\rm abs}\frac{N^{ij}}{N_{\rm min}^{ij}}
\end{equation}
keeping in mind that both $K_{\rm abs}$ and $N_{\rm min}^{ij}$ are orbit - dependent values.
Figure~\ref{fig:Flat_field_V4_Lightcurves} shows the raw, relatively-calibrated, and final flat-fielded lightcurves of two neighbouring pixels, and 
the same D3 frame is shown before and after flat fielding in Figure~\ref{fig:Flat_field_pacific_ocean}.
 
\begin{figure}[h]\centering
\centering
\includegraphics[width=.37\textwidth]{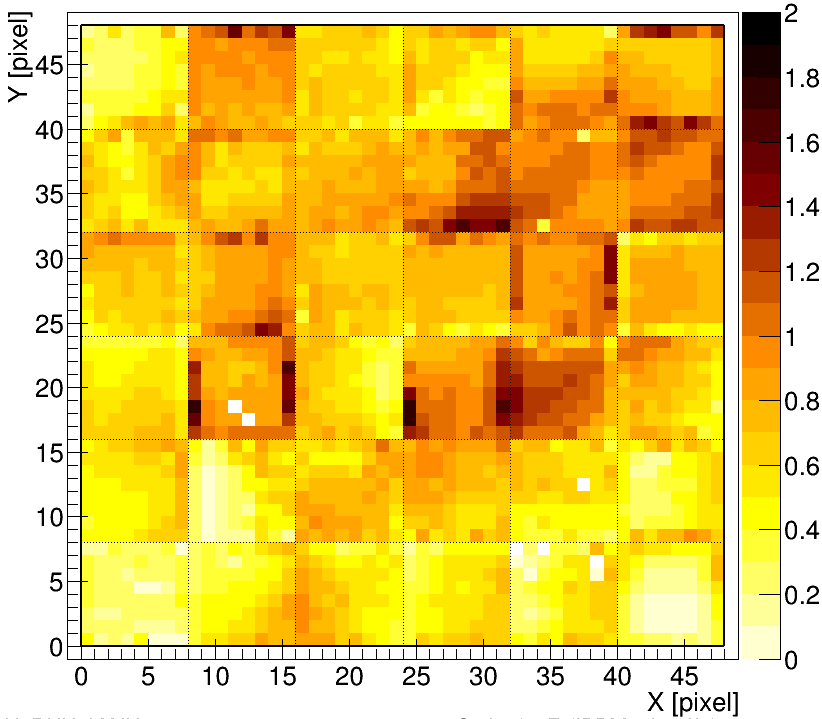} 
\includegraphics[width=.61\textwidth]{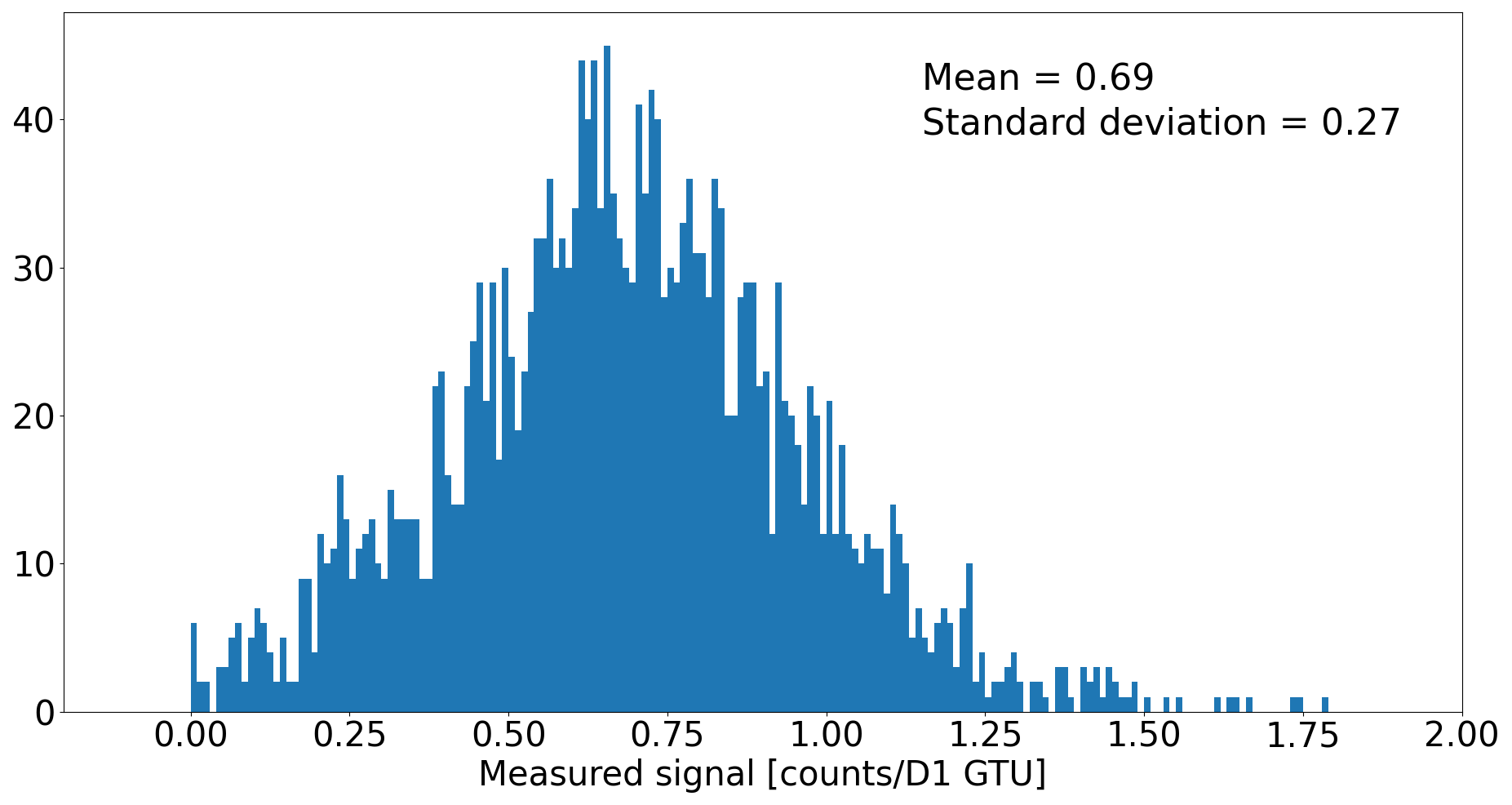} 
\includegraphics[width=.37\textwidth]{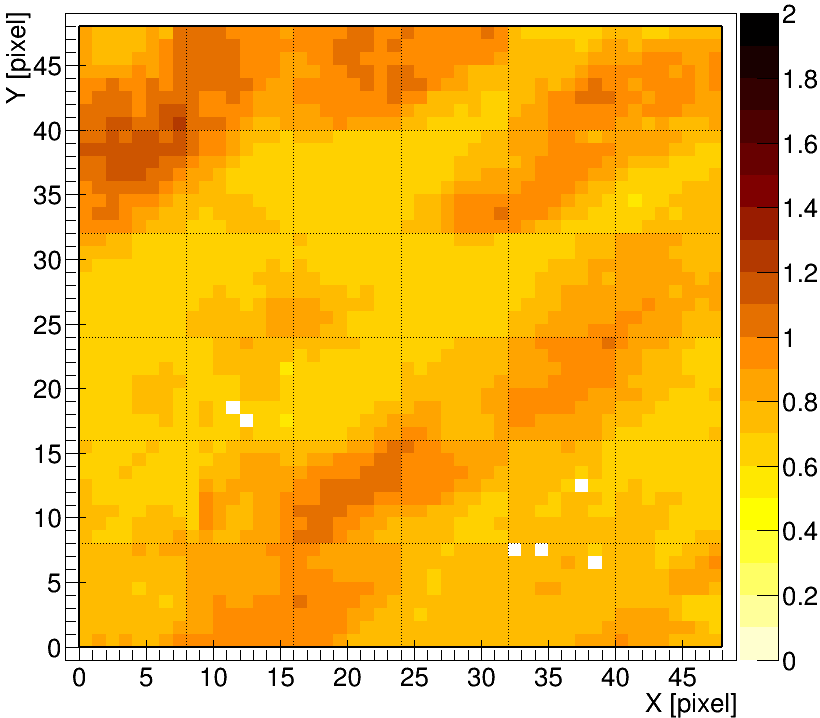} 
\includegraphics[width=.61\textwidth]{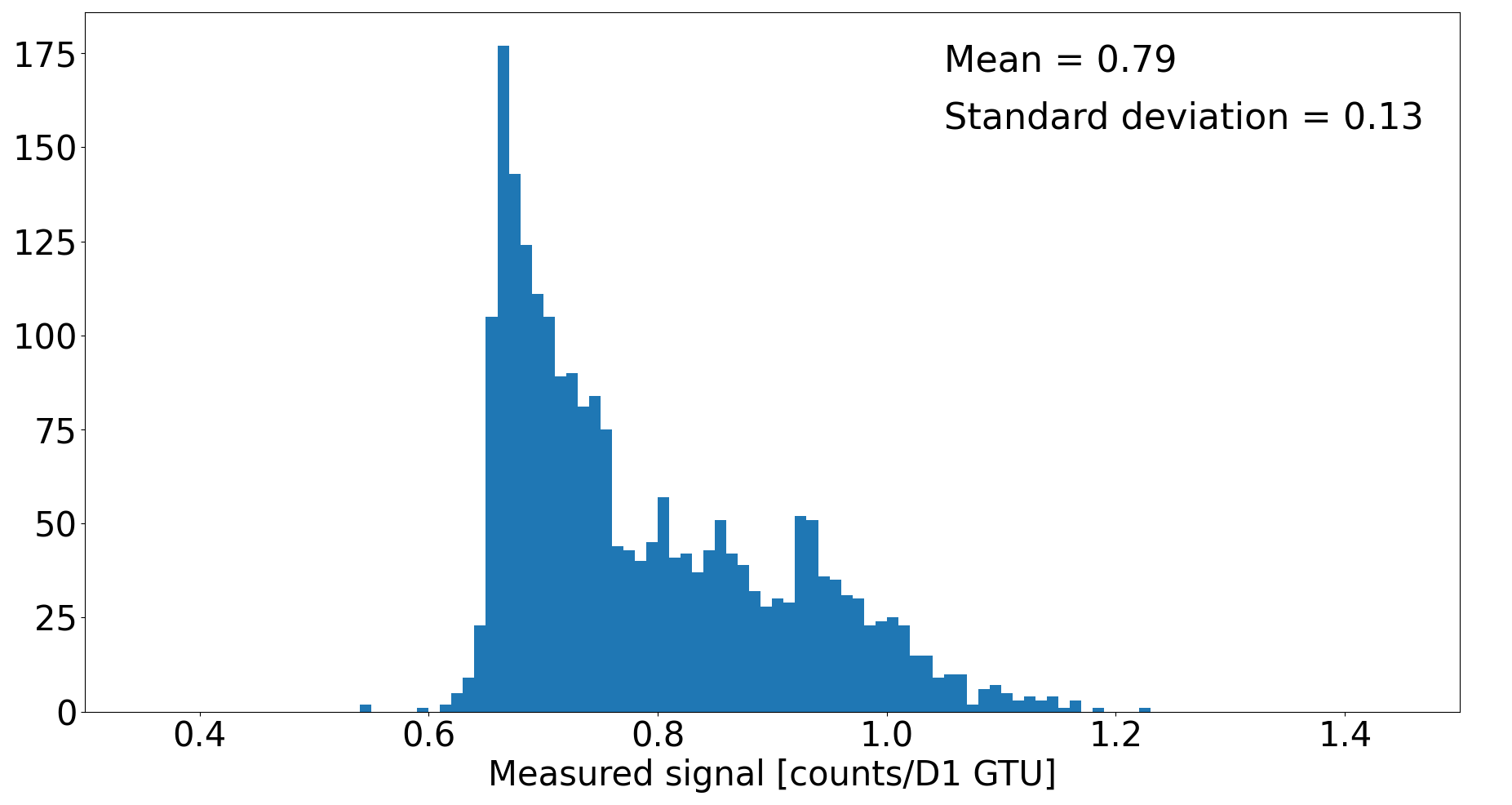} 
\caption{Response of Mini-EUSO in D3 (40.96~ms) timescale. \textbf{Top Left:} average of 128 raw D3 frames (a total of $\simeq$ ~5.1~seconds, normalized to  one GTU) of the sea off the coast of England. Also in this case, the different counts in each pixel are mainly due to the different efficiencies of the pixels. 
\textbf{Top Right:} histogram of the counts registered in the 2304 pixels.
\textbf{Bottom Left:} the same data shown above, after the flat fielding procedure. Details like the slightly more luminous bands produced by the clouds, are visible. \textbf{Bottom Right:} histogram of the flat-fielded data. Also in this case the spread is significantly reduced, with the tail above $\simeq 0.7$~counts/GTU  mostly due to the presence of clouds.  
 }
\label{fig:Flat_field_celtic_ocean}
\end{figure}

This acquisition comes from a cloudless and moonless passage over the Pacific Ocean. 
The image in the non flat-fielded picture is not uniform and a histogram of the  pixel counts shows a wide distribution with an average value of 0.56~counts/GTU and a standard deviation of 0.25~counts/GTU, with very long tales toward lower and higher values. 
The flat-fielded image, on the other hand, shows a much more uniform PDM, with an average value of 0.64~counts/GTU and a standard deviation of 0.12~counts/GTU. 
An example with clouds present is shown in Figure~\ref{fig:Flat_field_celtic_ocean}, taken when the ISS was flying over the Celtic Sea, off the coast of England.  
In this case, the raw image also exhibits a very wide distribution of counts, with artifacts present at the MAPMT borders.
After flat fielding, the distribution of the unclouded pixel is much sharper (peak at $\simeq 0.7$~counts/GTU), with the tail of high emissions (up to a factor $\simeq 2$ higher) due to the brighter diagonal cloud bands. 
 
\section{Sun and Moonlight}

In order to accurately estimate the UV emissions from Earth it is necessary to take into account  the contribution of reflected or stray light coming from extra-terrestrial objects, with the two dominant contributions being from the Sun and Moon. Light from bright planets and stars  can be neglected  since it does not contribute sensibly to the background. Thresholds for the Sun and Moon positions relative to the local horizon can be used to ensure that the light observed is coming from the Earth's surface and atmosphere.

\begin{figure}[h]
    \centering
    \includegraphics[width =1.\textwidth]{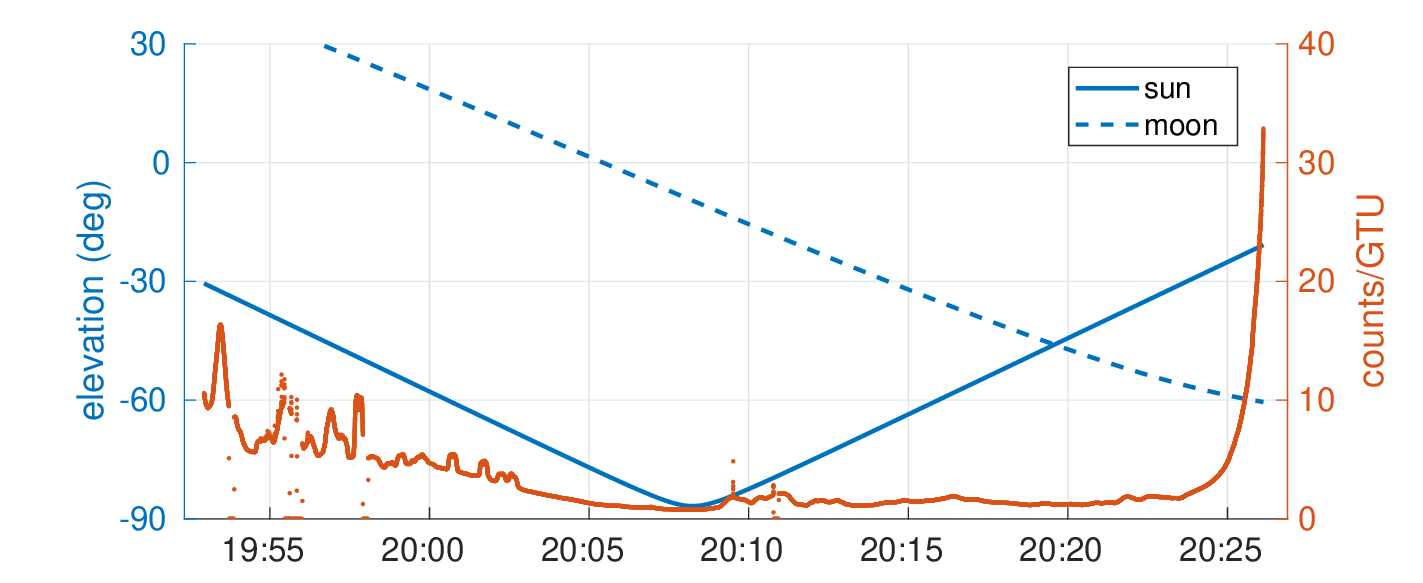}
    \caption{An example of counts/GTU averaged over the central  four  PMTs (orange) on the right y-axis and solar (solid blue) and lunar (dashed blue) elevation on the left y-axis. The x-axis indicates hours and minutes in UTC time of 2020-03-31, when the Moon was 44\% full. } 
    \label{fig:solar_elev_example}
\end{figure}

The mean number of counts/GTU over the central four PMTs of the detector is plotted as a function of time  in Figure~\ref{fig:solar_elev_example} for a typical acquisition run, lasting approximately one semi-orbit. A run  begins when the light - measured by a photodiode on the focal surface - falls below a safe threshold and ends as it passes above a second threshold. The Figure also shows the elevation of the Sun and Moon above the horizon, as seen from the ISS, calculated using the Skyfield \citep{SKYFIELD2019ascl.soft07024R} python package.  Towards the end of the passage, the number of counts increases due to scattered atmospheric light when the Sun is still below the horizon ($0^{\circ}$ on the left y-axis) but with elevation above $-30^{\circ}$. A more gradual modulation of the average detected signal  depends on the elevation of the Moon  (for this session the Moon phase was $m=0.44$) with an appreciable increase in the number of counts when the Moon is above the horizon and growing with its zenithal angle.

\begin{figure}[H]
    \centering
    \includegraphics[width = .7\textwidth]{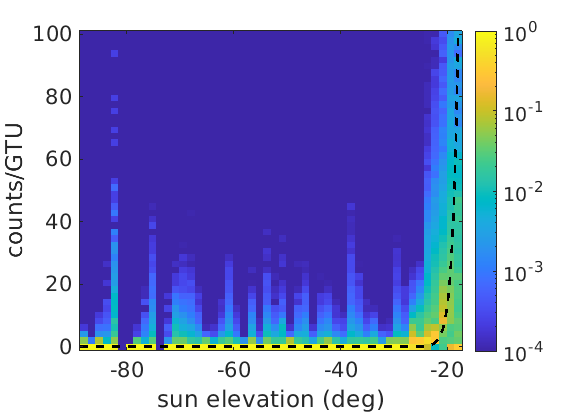}
    \caption{Effect of sunlight on Mini-EUSO: histogram of the distribution of the number of observations versus average counts/GTU (y-axis) and the solar elevation above the local horizon (x-axis). Only data with the Moon below the horizon, taken over sea, have been considered. Each solar elevation (a column of the histogram) has been normalized to one to take into account the different amounts of time spent at each solar elevations. The dashed line indicates a least squares fit to an exponential function.}
    \label{fig:sun_elev}
\end{figure}

A more detailed description of the effect of the Sun can be found in Figure~\ref{fig:sun_elev}. In this two-dimensional histogram, each bin has been filled according to the mean counts/GTU averaged over one pixel size (6.3$\times$6.3~km$^2$ areas) on the y-axis, and the elevation of the Sun on the x-axis. To take into account the different amounts of time spent at a given inclination, the observations for each solar elevation $\Theta_{\odot}$ (each column) have been normalized to 1, so that the color scale indicates how common a certain number of average counts is for a given elevation interval.  Observations when the Moon is above the horizon have been excluded to isolate the effect of the Sun. As in the previous Figure, a rapid increase of the light can be found for $ \Theta_{\odot}> -30^\circ$. This is confirmed by a least squares fit of the form $y = a\cdot e^{ \Theta_{\odot} - b}$, which  yielded $b = -29^\circ$, also shown in Figure~\ref{fig:sun_elev}. 

\begin{figure}[H]
    \centering
    \includegraphics[width = 0.45\textwidth]{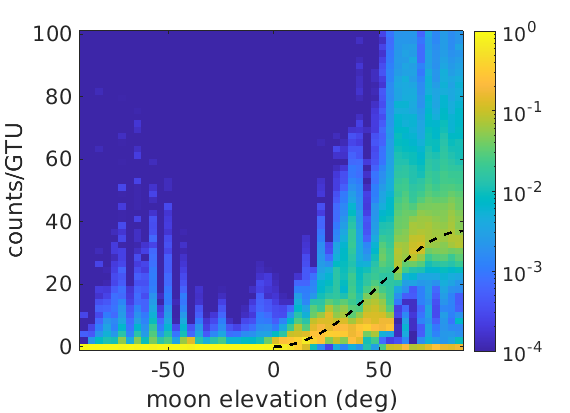}
    \includegraphics[width =0.45\textwidth]{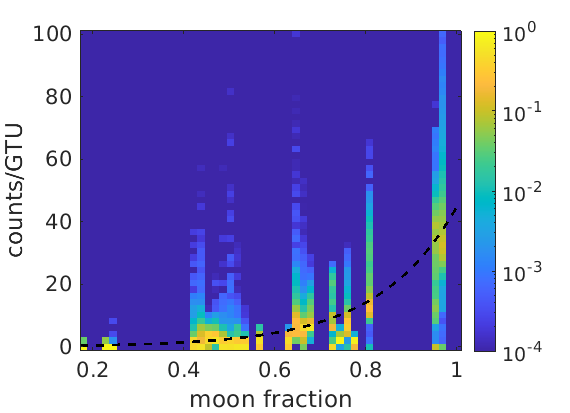}
    \caption{Effect of moonlight on Mini-EUSO: histograms of the relative distribution of observations for average counts/GTU (y-axis) as a function of the Moon elevation (left, including all Moon phases) and Moon fraction (right) on the x-axis. Dashed lines represent least square fits to equations \ref{eq:zenithAngle} (left) and \ref{eq:moonFracExponential} (right). Only data with the Sun less than 30$^{\circ}$ below the horizon, taken over sea and without clouds have been considered. For the right panel, data with the Moon below the horizon were discarded. Each Moon elevation and Moon fraction (columns of the two histograms) have been normalized to one to take into account the different time spent at the various lunar elevation and phases. }
    \label{fig:moon_elev_frac}
\end{figure}

Figure~\ref{fig:moon_elev_frac} shows similar plots for moonlight. The left panel of the figure shows the average light increase when the Moon is above the horizon; values of 40~counts/GTU and above are possible when the Moon is close to the zenith. The right panel shows the counts/GTU as a function of the Moon phase.  An increasing phase of the Moon appears to increase the counts exponentially, with higher average counts being observed when the Moon fraction is above 0.5, i.e.~greater than a half-moon.

The dashed line in the left panel of Figure~\ref{fig:moon_elev_frac} represents a fit according to the following formula \citep{Garipov2005}:

\begin{equation}\label{eq:zenithAngle}
    I(m,\theta,\lambda) = F(m,\lambda)\left[1-\exp\left(1-\frac{X_0(1+\cos\theta)}{X_R(\lambda)\cos\theta}\right)\right]\frac{\cos^2\theta(1+\cos^2\theta)}{3\pi(1+\cos\theta)}, 
\end{equation}

where $I(m,\theta,\lambda)$ is the back-scattered light intensity, $F(m,\lambda)$ is the moonlight intensity for a Moon fraction $m\in[0,1]$ and wavelength $\lambda$, $\theta\in[0^\circ,90^\circ]$ is the local zenith angle of the Moon, $X_0$ is the depth in the atmosphere open to observation and $X_R(\lambda)$ is the Rayleigh scattering length. $X_0 > X_R$ corresponds to cloud-free conditions, and the ratio $X_0/X_R$ was fixed to 2 to represent this   {condition}.  
Since the exact distribution of the convolution of emitted  wavelengths $\lambda$ and the Mini-EUSO efficiency is difficult to estimate,   $F(m,\lambda)$ is set as a free parameter, $F_{est}$, in the least squares fit. The resulting estimate was $F_{est} = 348.7 \pm 0.4$.

If we assume an exponential dependence of $F$ from the Moon phase $m$, data displayed in the right panel of Figure~\ref{fig:moon_elev_frac} can be fitted according to:   \begin{equation}\label{eq:moonFracExponential}
    F(m) = ae^{b\cdot m}.
\end{equation}
The least squares estimates of these parameters were $a  = 0.632 \pm 0.003$~counts/GTU and $b  = 4.036 \pm 0.006$. While these models are not further used in this paper, they could potentially be helpful when interpreting data collected in moonlight conditions.

\begin{figure}[H]
    \centering
    \includegraphics[width = 0.42\textwidth, height = 0.41\textwidth]{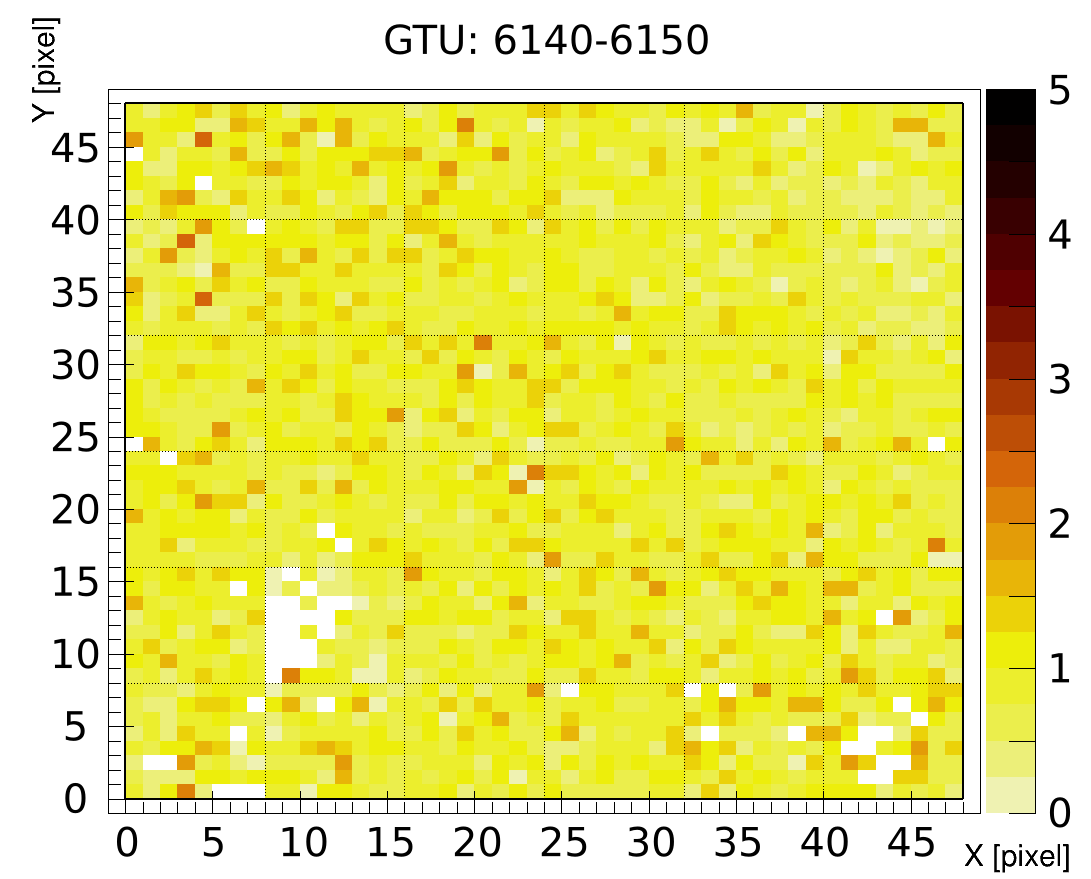}
    \includegraphics[width = 0.42\textwidth, height = 0.41\textwidth]{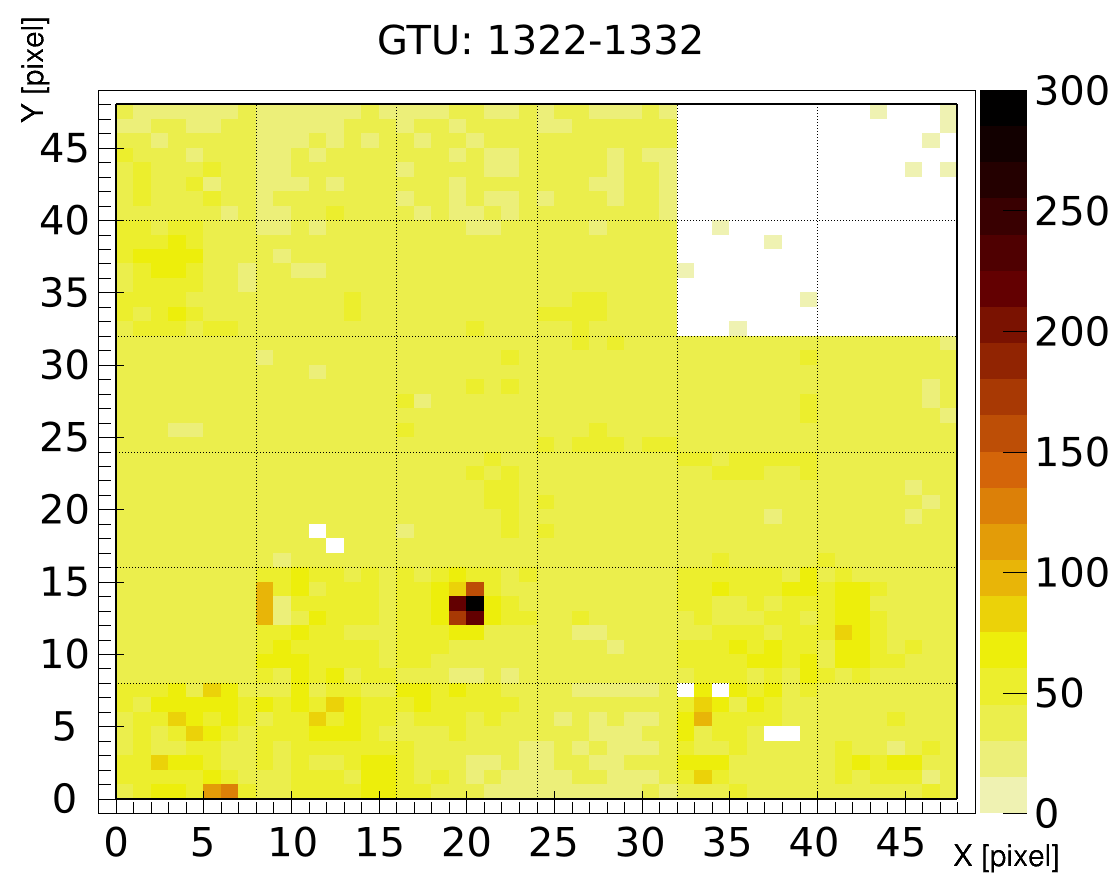}
    \caption{D3 counts averaged over 10 frames ($\simeq$~410~ms) when passing over areas of the Sahara desert (centre of the PDM  being at (17.1$^\circ$,1.8$^\circ$), and (15.6$^\circ$, -4.6$^\circ$) in the left and right panels respectively). In the left panel the Moon is below the horizon, while in the right panel it has an elevation of 79.0$^\circ$ and a fraction of 0.95, resulting in a direct reflection off of the desert. Given its distance from the Earth, the Moon appears not to move in PDM coordinates, so that the light can be summed over several frames. }
    \label{fig:sahara_moon_examples}
\end{figure}

When the Moon gets close to the zenith, the light reflected from the sea to Mini-EUSO is too high and causes the  EC safety mechanism to lower the high voltage.  However, when this condition occurs above the ground - where the albedo is lower  - it is possible to see the direct reflection of the Moon. An example of this is illustrated in Figure~\ref{fig:sahara_moon_examples}, where the left panel \textcolor{black}{(taken on 2021-05-06, 22:16:10 UTC)}  shows a 10 D3 frame averaged image taken over the Sahara desert with the Moon below the horizon, and the right panel \textcolor{black}{(taken on 2020-01-08, 08:22:03 UTC)}  shows a direct reflection of the near-full Moon ($11^\circ$ from zenith) off of the surface of the desert. The average signal over 10~frames ($\simeq$~410~ms) is $\simeq 158$~counts/GTU for the four center-most pixels. Note that the Moon light is fixed in PDM coordinates, but the ground upon which it is reflected is moving so that the mean counts represent an average of the albedo of the terrain where it is reflected.

\section{Cloud coverage}\label{sec:clouds}

As for the back-scattered light from the Sun and Moon, {in order} to study the UV light emission from ground detected by Mini-EUSO, it is important to know the cloud coverage below the telescope. To this purpose, information from a numerical  weather prediction model is used,  in particular, data from the US National Weather Service Global Forecast System  \citep[GFS][]{GFS_cisl_rda_ds084.6}. 
    {This information gives real-time and completely automated forecasts of the cloud coverage}. Global models like GFS discretize the Earth's atmosphere in cells with a vertical distribution of levels (in this case 57 from ground to \SI{1}{\hecto\pascal}, roughly \SI{80}{\kilo\metre} above the ground level), and a horizontal latitude-longitude grid, spaced at \ang{0.25} (\SI{28}{\kilo\metre} at the equator).

The model predicts surface variables, such as temperature, wind speed, and surface turbulent fluxes, and also evaluates the atmosphere dynamics and thermodynamics on the entire atmosphere volume considered, calculating winds, humidity, temperature, and cloud cover fraction. 
The last of these is the most useful variable in this context,   {and} on each model level we have the cloud fraction coverage (as a percentage, where 0\% is clear sky and 100\% is overcast)   which we aggregate  into  three standard cloud heights: low (from ground to \SI{2}{\kilo\metre} above sea level (a.s.l.)), medium (\SIrange{2}{6}{\kilo\metre} a.s.l.) and high (\SIrange{6}{18}{\kilo\metre} a.s.l.) clouds. Above the tropopause, located between \SIrange{8}{18}{\kilo\metre} a.s.l.~from the poles to equator, clouds are rare and do not influence the observations conducted with the Mini-EUSO detector.

The GFS calculates a new forecast every 6 hours at the standard meteorological times: 00:00, 06:00, 12:00, and 18:00~UTC. 
  {  At these  analysis times,  } all the meteorological and surface data are assimilated and interpolated on the model grid,   {and from this starting point} the model starts the forecast with a time-step of 1 hour. 
The evolution of the weather between two forecast steps is not available, as it is only computed inside the model.
Further analysis could be done on specific detected events, considering observations (such as satellite images,   {if available}, atmospheric soundings,   {  etc.}) or with high-resolution regional models, as done by \citep{Khrenov2021}.

\begin{figure}[htbp]
     \centering
    \includegraphics[width = 1.1\textwidth]{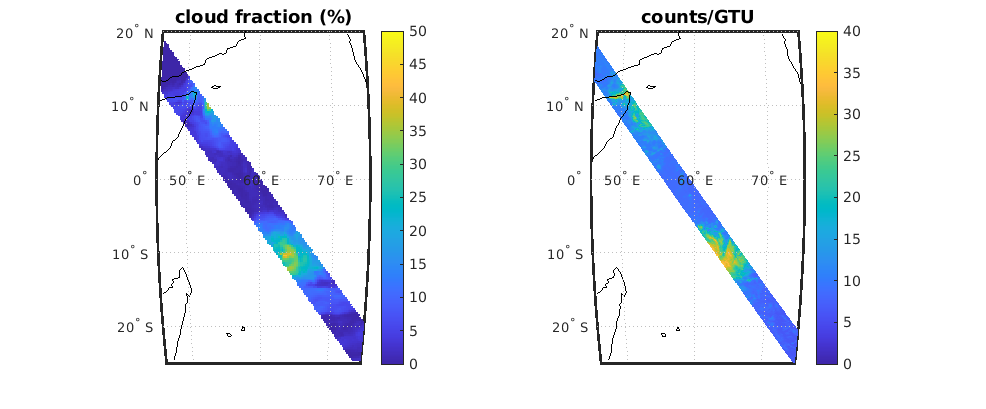}
    \includegraphics[width = 1.1\textwidth]{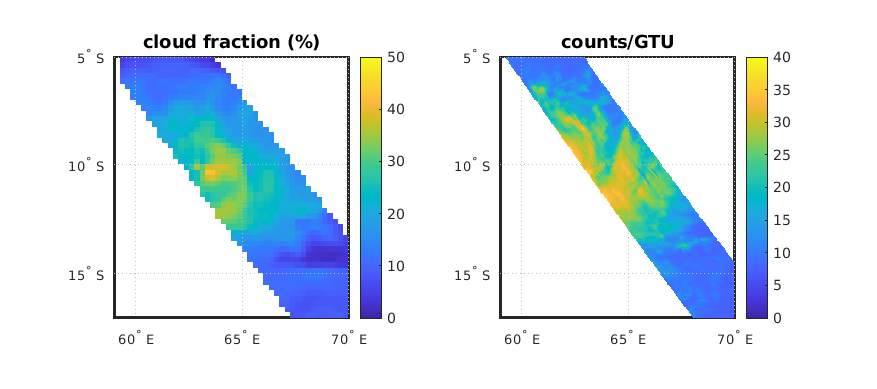}
     \caption{\textbf{Top:}   Comparison of cloud images from GFS (left) and from Mini-EUSO (right) for a passage starting on 2019-12-05, 18:30 UTC. A good agreement between the modelled cloud cover and the observation by Mini-EUSO is visible.
     \textbf{Bottom:} a close up of the above passage showing a cyclonic region with its center outside the Mini-EUSO field of view. Mini-EUSO UV counts in $0.05^\circ \times~0.05^\circ$ map cells (right), and GFS cloud fraction (\%) with $0.25^\circ \times~ 0.25^\circ$ of spatial resolution (left) on 2019-12-05, 18:38 UTC. Note that the high phase (0.64) and elevation ($33.9^\circ$ above the horizon) of the Moon gave rise here to a high number of counts/GTU.}
     \label{fig:cloud_map_comparison}
 \end{figure}

\begin{figure}[htb]
     \centering
     \includegraphics[width = .45\textwidth]{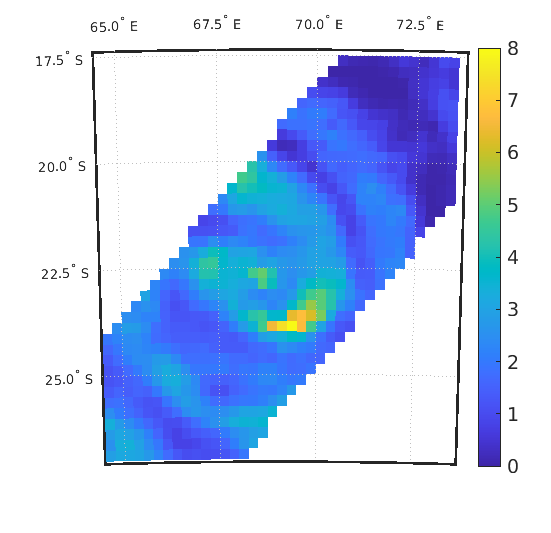}
     \includegraphics[width = .45\textwidth]{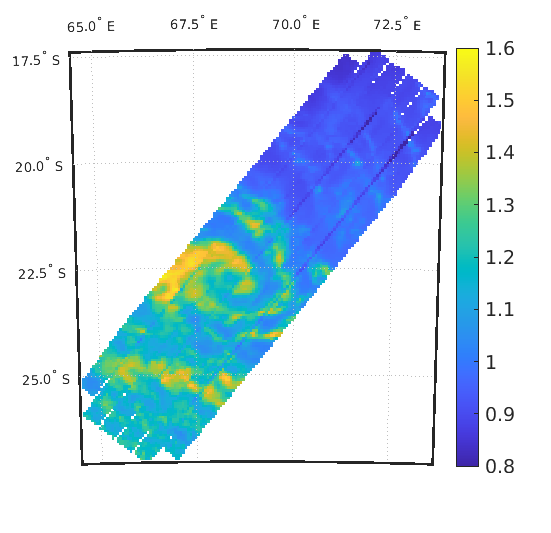}
     \caption{Modelled cloud fraction (\%) with $0.25^\circ \times~ 0.25^\circ$ of spatial resolution (left) and Mini-EUSO UV counts in $0.05^\circ \times~ 0.05^\circ$ map cells (right). Acquisition on 2020-02-21, 22:00~UTC, $\simeq 1000$~km East of Mauritius island. In this case the Moon was below the horizon ($-27.6^\circ$), and therefore the cloud brightness is about 1.6~counts.}
     \label{fig:small_swirl}
\end{figure}

In Figure~\ref{fig:cloud_map_comparison},  a comparison between the cloud fraction from GFS\footnote{The model used as reference during this session is the ERA 5 reanalysis \citep{ERA5} because the real-time GFS forecast was missed. The discrepancies between the two models are modest.} (left panels) and from Mini-EUSO (right panels) is shown along the path of an  orbit (top panels) and in a cyclonic region (bottom panels). The figure shows the good agreement between the modelled and the observed cloud cover and the good detector resolution which allows us to see the structure of the cyclone.
Figure \ref{fig:small_swirl} shows a similar comparison between the modelled cloud distribution and the Mini-EUSO counts,  with the Moon below the horizon: also in this case the cyclonic structure is visible and is in good agreement with the modelled cover, albeit with a much fainter signal  ($\simeq 40$~counts/GTU vs $1.6$~counts/GTU, a ratio of $\simeq 25$) than in the previous example.

\begin{figure}[hbt]
    \centering
     \includegraphics[width =1.0\textwidth]{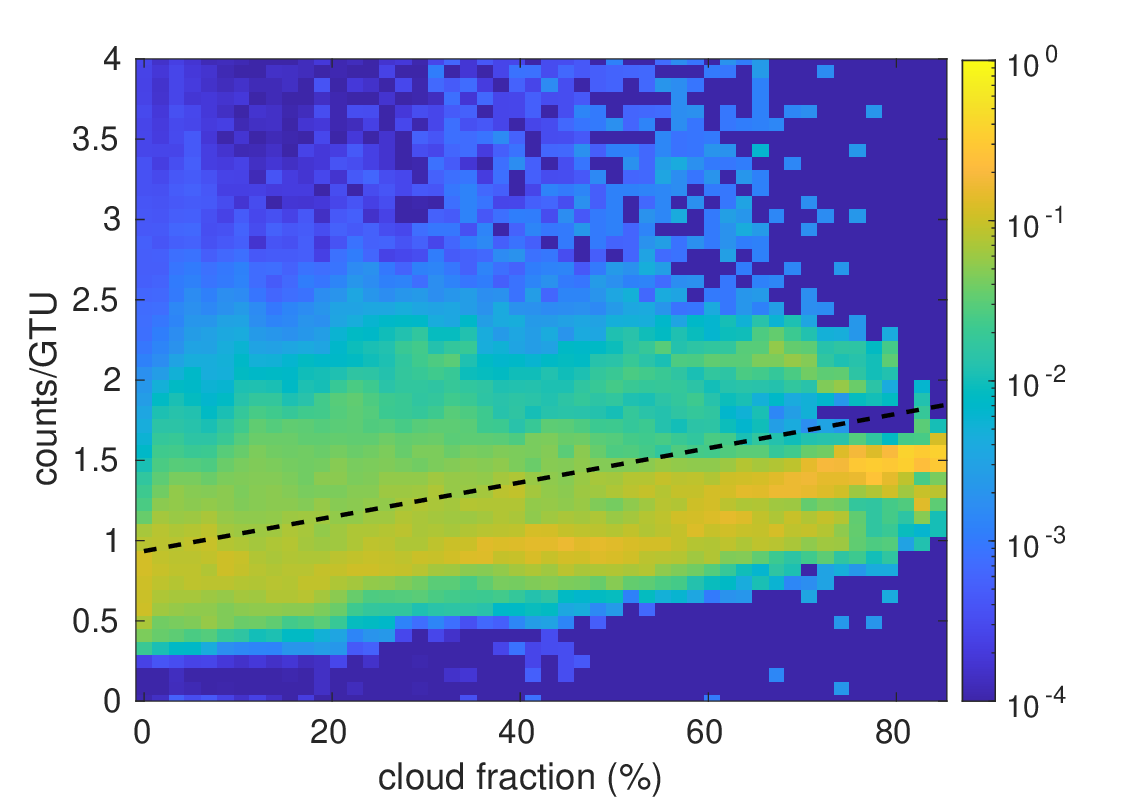}
    \caption{Histogram of the distribution of the percentage of observation time (z-axis) as a function of the average counts/GTU (y-axis) and the GFS cloud fraction (x-axis). Only data over  the sea with both the Sun and Moon below the horizon have been considered. Each cloud fraction interval (a column of the histogram) has been normalized to 1 to take into account the different time spent at the various cloud conditions. The dashed line represents a linear fit of the dependence of counts - cloud fraction.  }.  \label{fig:cloud_hist_and_fit}
\end{figure}

Figure~\ref{fig:cloud_hist_and_fit} shows a linear regression $y = c_0 + c_1x$ for counts/GTU in the y-axis against the cloud fraction on the x-axis for sea areas, with the Moon (and Sun) below the horizon. The resulting estimate are  $c_0=0.93\pm 0.01$~counts/GTU  and $c_1=0.011\pm 0.002$~counts/(GTU$\cdot\%$ cloud cover), i.e.~about 0.1~counts/GTU for each additional 10\% of cloud cover.

\section{Emissions from the ground and the atmosphere}

As discussed above, Mini-EUSO detects photoelectron counts coming from sources of various origin, shape, and duration. The amount of light detected by Mini-EUSO depends on the emission spectrum, the size (e.g.~point-like or diffuse), and the altitude of each source. We can divide the events observed with Mini-EUSO into different classes, according to their nature:  
\begin{enumerate}
    \item {\bf Point-like sources on the ground}. These include small villages, single fishing boats, Xenon flashers, etc. In this case  we can assume a negligible spatial dimension of the source. If we assume an isotropic emission, the photon flux  $\Phi$ is:
    \begin{equation}
        \Phi = C \cdot \frac{1}{\epsilon(\lambda)} \frac{2\pi H^2}{A} \frac{1}{\textrm{Atm}(\lambda)}
    \end{equation}
    where  $C$ is the number of counts per unit time, $\epsilon(\lambda)$  is the  detection efficiency of Mini-EUSO ($\epsilon_{\rm average}=0.080\pm 0.015$), $H$ is the distance of the ISS from the source (e.g.~if the source is on the ground at sea level $H=4\cdot 10^5~$m), $A=0.05~$m$^2$ is the area of the entrance pupil of Mini-EUSO and  $\textrm{Atm}(\lambda)$  is the atmospheric absorption coefficient (with $\textrm{Atm}_{\rm average}=0.74$). Since the efficiency and the atmospheric absorption depend on the wavelength $\lambda$, calculating the number of photons emitted by a  light source requires a precise knowledge of the emission spectrum of the source, or the best possible approximation. 
    For instance, a ground source, emitting isotropically on a half sphere at 395~nm and producing $C$ photoelectron counts/GTU in Mini-EUSO, emits on the ground: 
    \begin{equation}
        \begin{aligned}
        \Phi ({\rm photons/ns}) & = C~{\rm pe/GTU} \frac{1}{2500~{\rm ns/GTU}} \frac{1}{0.08}{\rm ph/pe} \frac{2\pi \cdot16\cdot 10^{10}\: {\rm m^2}}{5\cdot 10^{-2}~{\rm m^2}} \frac{1}{0.74} \\ 
        & = C\cdot (1.4\pm 0.3) \cdot 10^{11}~{\rm ph/ns}
        \end{aligned}
    \end{equation}
    The power emitted  by a 395~nm source which results in 1 photoelectron count being detected in Mini-EUSO (assuming a 400~km altitude of the ISS) is:
    \begin{equation}
        P  ({\rm W})= \Phi ({\rm photons/s})  \frac{hc}{\lambda} = 1.4\cdot 10^{11}~{\rm ph/ns} \cdot 10^9 ~{\rm ns/s} \cdot 5.03\cdot 10^{-19}~{\rm J/ph}  = 70\pm 15~{\rm W}
    \end{equation}
  This value can be assumed as an estimation of the minimum transient (flashing within one GTU) point-like source  detectable by Mini-EUSO in absence of background emissions. This value can be lower for constant signals acquired in  D2 and D3 mode, where the signal/background ratio is higher.

     \item  {\bf Diffuse sources on the ground}, with an extension comparable or larger than one pixel ($6.3 \times 6.3~$km$^2$  on the ground for a ISS altitude of 400~km). These include  large towns, fleets of fishing boats, and sea bioluminescence. In this case, we can assume that the light is uniformly distributed over a wide area. Here the flux $\phi$ for a telescope with entrance pupil $A$ (as above) with a pixel observing a field of view of $\Omega$ is:
    \begin{equation}
        \phi = C \cdot \frac{1}{\epsilon(\lambda)} \frac{ 1}{\Omega} \frac{1}{A}\frac{1}{\textrm{Atm}(\lambda)}
    \end{equation}
 where 
    \begin{equation}
    \Omega= \left( 2\cdot \tan\frac{\theta}{2}\right)^2 = \frac{ L^2}{H^2} =  2.48 \cdot 10^{-4}~{\rm sr}
    \end{equation}
     and $\theta=0.9^{\circ}$ is the field of view of one pixel, $L\simeq 6.3$~km is the side of the - in first approximation square - area observed by one pixel.
     In our case, assuming a diffuse source at 395 nm, the flux is: 
\begin{equation}
    \begin{aligned}
        \phi ({\rm photons/(ns~m}^2{\rm~sr})) &= C~{\rm pe/GTU} \frac{1}{2500~{\rm ns/GTU}}  \frac{1}{0.08} {\rm ph/pe} \frac{1}{2.48 \cdot 10^{-4}~{\rm sr}}  \frac{1}{5\cdot 10^{-2}~{\rm m^2}}  \frac{1}{0.74}  \\
        &= C\cdot(550 \pm 100) ~{\rm ph/(ns~m}^2~{\rm sr})
    \end{aligned}
\end{equation}

    \item {\bf  Diffuse sources in the upper atmosphere}.  These include  night-time airglow, the main source of UV emission over the Earth's surface,  Transient Luminous Events (TLE), such as ELVES, and aurorae (although the latter are rarely visible at the ISS orbit).  For these events there is no atmospheric absorption so that:
    
    \begin{equation}
        \phi_{\rm atm} ({\rm photons/ ( ns~m}^2~{\rm sr}) )=  C \cdot \frac{1}{\epsilon(\lambda)} \frac{ 1}{\Omega} \frac{ 1}{A}  = C\cdot (400 \pm 80) ~{\rm ph/( ns~m}^2~{\rm sr)}
    \end{equation}
    this is valid mostly for transient phenomena such as ELVES where the ground reflection can be neglected\footnote{Here we mean the reflected light, which will arrive $\Delta t= 2\times (90 ~$km$) / (3\times 10^5 ~$km/s$) \simeq 600~\mu$s later.}. In case of constant uniform emissions, if we want to assess the brightness of airglow at the emission point,  the flux is:
    \begin{equation}
        \phi_{a\rm tm} ({\rm photons/( ns~m^2~sr)})=  C \cdot \frac{1}{\epsilon(\lambda)} \frac{ 1}{\Omega} \frac{ 1}{A} \left(  1+ \frac{  r}{\textrm{Atm}(\lambda)\cdot \textrm{Atm}(\lambda)} \right) 
    \end{equation}
    where $r$ is the albedo of the ground, and we have to take into account light travelling down to the surface of the Earth  and then up again to the detector. 
    Earth UV albedo in the 300 - 400~nm range is about 0.3 \citep{MEIER_UV_1991SSRv...58....1M}.  
    Assuming $r=0.3$ we have a correction factor of 1.55, resulting in a flux of  $C \cdot ( 620\pm 120)$~ph/(ns~m$^2$~sr).

    \item  {\bf Clouds}. These can be considered as diffuse structures that reflect the atmospheric glow, starligh, and moonlight with a higher efficiency than the ground.  
     Neglecting atmospheric absorption,  the calibration factor is the same as for the airglow. If we assume that the higher brightness of the cloud is due to reflection of the airglow above it, we can estimate the reflection factor of the cloud (related to its optical thickness) as:
     
     \begin{equation}
        r_{\rm cloud} = \frac{\phi_{\rm cloud}}{  \phi_{\rm atm}}   
    \end{equation}

\end{enumerate}

\section{Earth Observations and Mapping}

\begin{figure}[H]
    \centering
    \includegraphics[width = 0.9\textwidth]{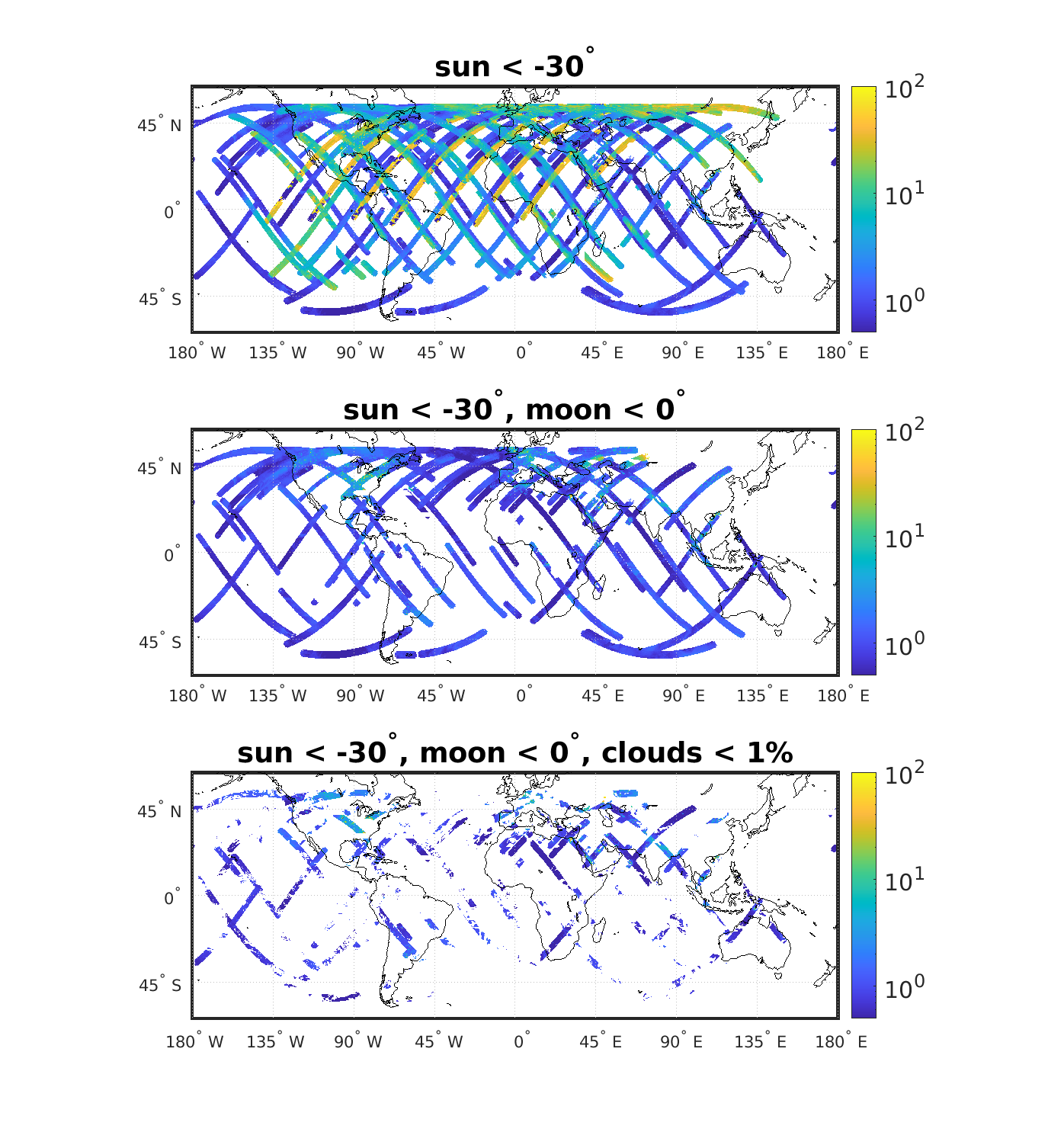}
    \caption{Mini-EUSO data mapped onto the surface of the Earth in $0.1^\circ\times0.1^\circ$ spatial bins. \textbf{Top:} observations with the Sun position more than 30$^{\circ}$ below the local horizon. \textbf{Middle:} observations with Moon position below the local horizon in addition to the previous Sun position selection criteria. \textbf{Bottom:} observations with Moon position below the local horizon and cloud fraction below 1\% in addition to the Sun position selection criteria. The observations refer to data acquired between 2019/11/19 and 2021/05/06 (32 sessions). \textcolor{black}{ The study of medium to long term variations in the ground emission will be the subject of future papers, with a larger dataset.}
   \textcolor{black}{Eventually, it is expected that Mini-EUSO will be able to survey the entire Earth surface in the latitude range $\pm 51.6^{\circ}$ (the orbital inclination of the ISS). However, the Pacific regions are sampled with a lower frequency, since the detector is turned on in the beginning of the crew night-time  of the station, at about 06:30~UTC. This corresponds e.g. to 05:30~AM on the local time in Japan, thus resulting in a more limited observation time before local dawn.}\label{fig:global_maps} }
\end{figure}

To produce global night-time UV maps of the Earth, we have  binned the data into 0.1$^\circ$ $\times$ 0.1$^\circ$ cells, showing the average number of counts per Mini-EUSO pixel normalized to one (D1) GTU. We have binned each point on the ground with the time-normalized counts of all the  pixels that were observing it during a passage. 
The  top panel of Figure~\ref{fig:global_maps} shows a subset of the available Mini-EUSO data, further selected for the regions where the Sun was below the horizon with an elevation $< -30^\circ$.  Most of the diffuse bright regions are due to cloudy areas with the Moon over the horizon. The central panel of the same figure shows the regions where the Moon was not visible (elevation threshold of 0$^\circ$). In the case of ground observations, a further cut  can be applied requiring the cloud fraction to be below a certain threshold: the bottom panel of the same picture shows the regions with a selection of less than $1\%$ cloud cover.   

\begin{figure}[H]
     \centering
   
    \includegraphics[width = 1.05\textwidth]{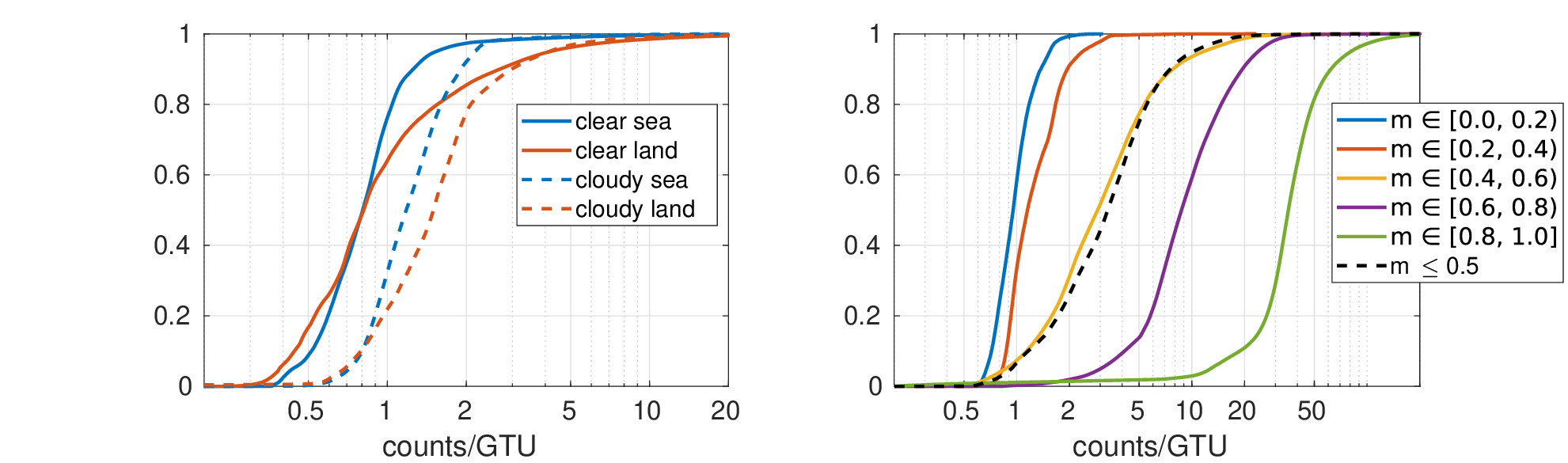}
     \caption{\textbf{Left:} cumulative distributions of Mini-EUSO observations over land and sea regions. Solid lines indicate moonless and cloudless conditions. Dashed lines represent moonless and cloudy conditions, i.e.~GFS cloud fraction above 1\%. \textbf{Right:} cumulative distributions of observations with the Moon visible above the horizon, each curve representing a different moon-phase interval.} 
     \label{fig:hist_land_sea_cdf}
 \end{figure}

\begin{figure}[H]
     \centering
     \includegraphics[width = 0.6\textwidth]{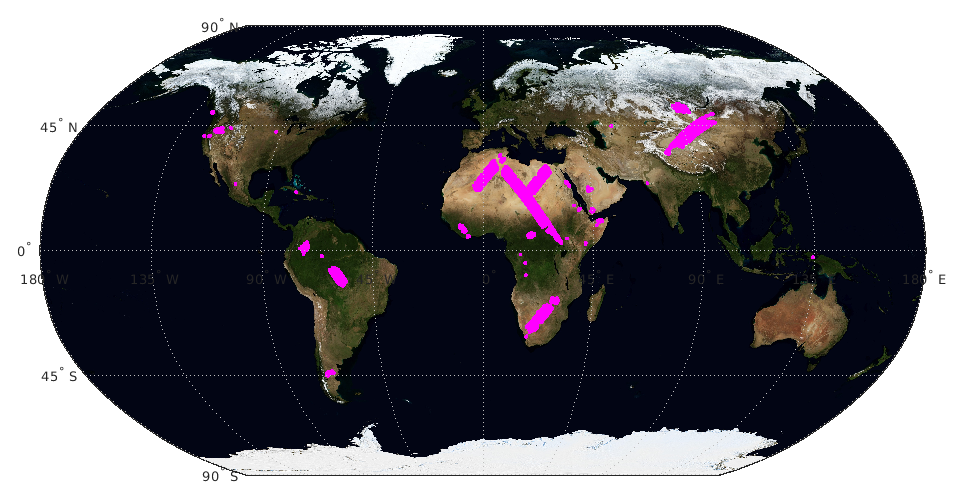}
     \caption{Locations (in magenta) over land with average counts/GTU below 0.5 in cloudless, moonless, conditions.}
     \label{fig:deserts}
 \end{figure}

Among the moonless observations, 32\% of them are estimated to be free from clouds (cloud fraction \textless 1\%). The majority of these are over the oceans (67\% in good agreement with Earth's sea/land ratio of 71\%). 
In Table \ref{tab:table_moon_cloud} are shown some typical values of the counts observed in various cloud and sea/land conditions.  
Sea emissions tend to produce lower average counts than measurements taken over land, as also seen in Figure~\ref{fig:hist_land_sea_cdf}. This Figure also shows that $\simeq 16\%$ of land areas have values below 0.5~counts/GTU, in comparison with $9\%$ of areas covering bodies of water. These areas of very low brightness are mainly distributed over the Amazon rainforest and the Sahara, Kalahari, and Taklamakan deserts (see Figure \ref{fig:deserts}).

 \begin{table}[!ht]
     \centering
    \begin{tabular}{|c|cc|cc|cc|cc|cc|}
         \hline
         & clear sea  & & clear land&  & cloudy sea &  & cloudy land &  & cloudy all  &\\\hline
         &   mean & STD  &   mean & STD  &    mean  & STD  &    mean  & STD  &   mean  & STD  \\\hline
         No Moon   &    0.9 &  0.4 &    1.4 &  1.6 &     1.3 &  0.4 &     1.7 &  1.1 &      1.4 & 0.7 \\
         Half Moon &      1.8 &  0.7 &      2.6 &  2.8 &     13.0 &  8.6 &      8.1 &  3.8 &      9.7 & 6.2 \\
         Full Moon &    37.6 & 10.9 &   35.1 & 10.4 &   50.7 & 19.7 &    51.1 & 18.6 &    51.0 & 18.6\\\hline
     \end{tabular}
     \caption{Average emission values  {(counts/GTU)} and their dispersion for sea and ground for various lunar phases and cloudiness. Half Moon includes moon fractions between 0.4 and 0.5, and full Moon includes fractions between 0.9 and 1. The brightest pixels (above 99th percentile) were excluded when calculating the mean and standard deviation to mitigate the effects from bright anthropogenic sources. For conditions with multi-modal distributions, the mode closest to the average is displayed.     } 
      
     \label{tab:table_moon_cloud}
 \end{table}

\begin{figure}[H]
    \centering
    \includegraphics[width = .9\textwidth]{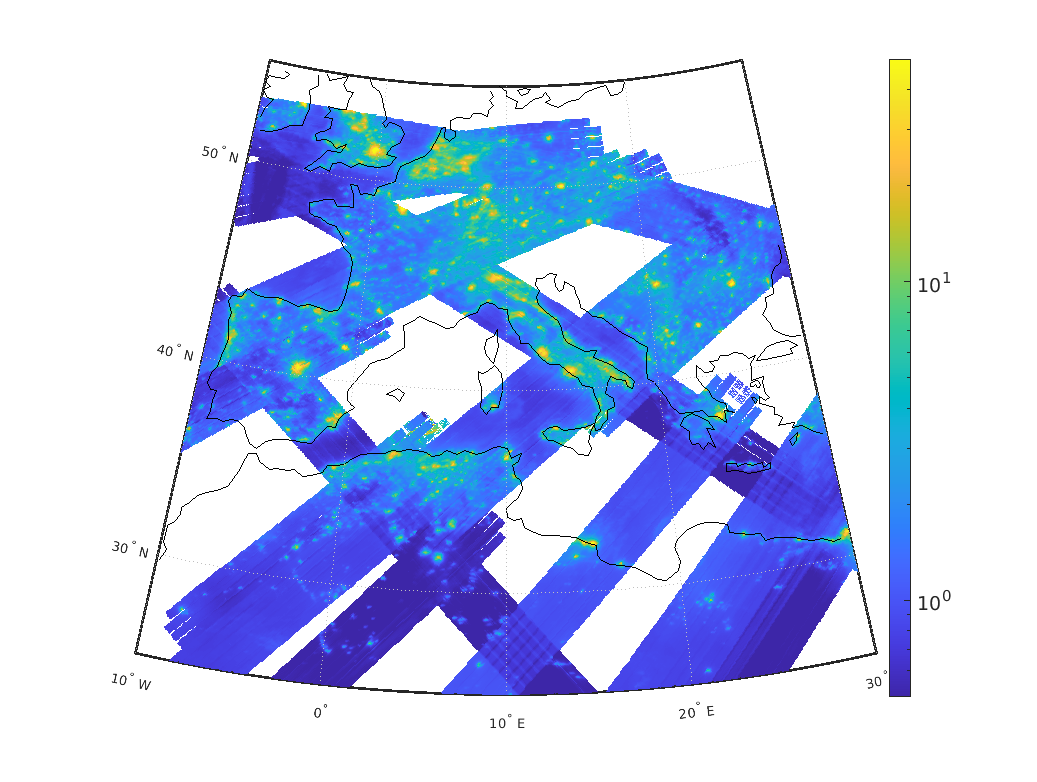}
    \includegraphics[width = .9\textwidth]{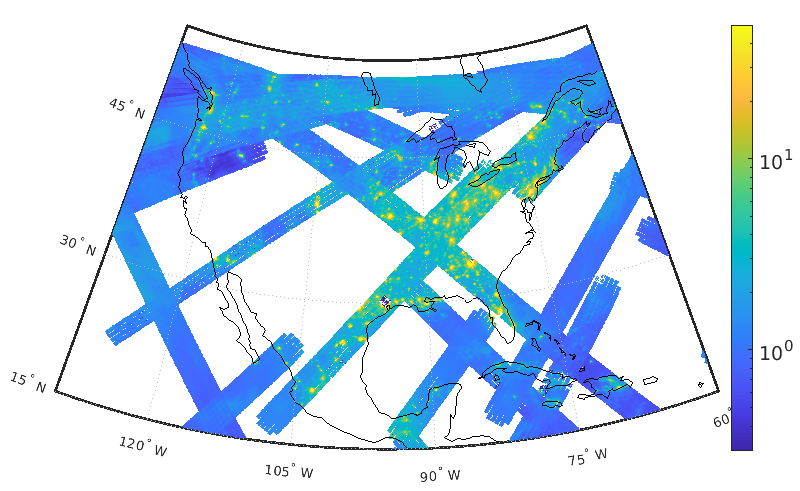}
    \caption{\textbf{Top:} counts/GTU in moonless conditions over parts of Europe and North Africa. Note the relative darkness in areas over sparsely populated areas like the Sahara desert and the Carpathian and Apennine mountains. \textbf{Bottom:} counts/GTU in moonless conditions over North America.}
    \label{fig:europe_map}
\end{figure}

\begin{figure}[H]
     \centering
     \includegraphics[width = 0.45\textwidth]{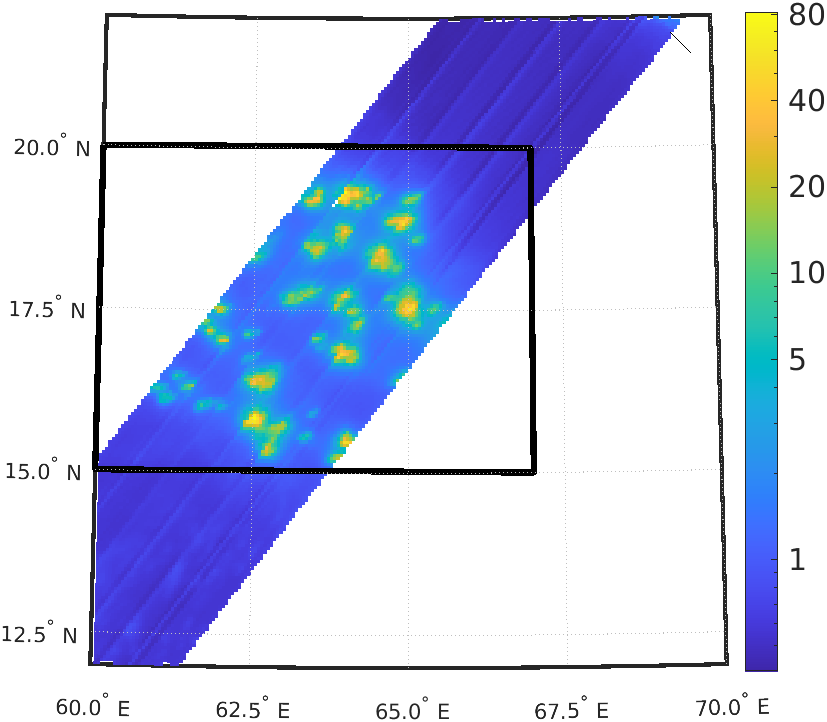}
    \includegraphics[width = 0.45\textwidth]{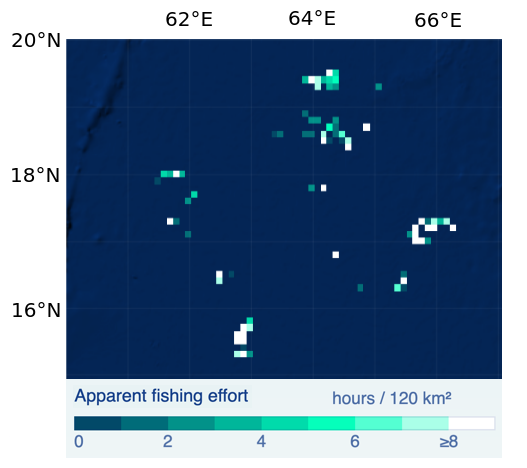}
     \caption{\textbf{Left:} average UV counts/GTU in $0.05^\circ\times~ 0.05^\circ$ map cells. \textbf{Right:} map of apparent fishing activity from the Global Fishing Watch [\url{https://globalfishingwatch.org}] corresponding to the area indicated in the left panel. The maps in both panels cover an area in the Arabian Sea on 2020-03-03.}
     \label{fig:boats}
 \end{figure}

 More detailed maps are shown in Figure~\ref{fig:europe_map}:

they reveal the impact of population density on the amount of UV light detected by Mini-EUSO, which is much more dramatic than the increase of the counts due to e.g.~atmospheric conditions.   Although most of the artificial light sources are over ground, regions with high counts/GTU over the ocean are due to fishing boats, where the light is  used to draw fish to the nets and is reflected on the surface of the sea (see Figure~\ref{fig:boats}). 

\begin{figure}[htb]
    \centering
    \includegraphics[width = 0.5\textwidth]{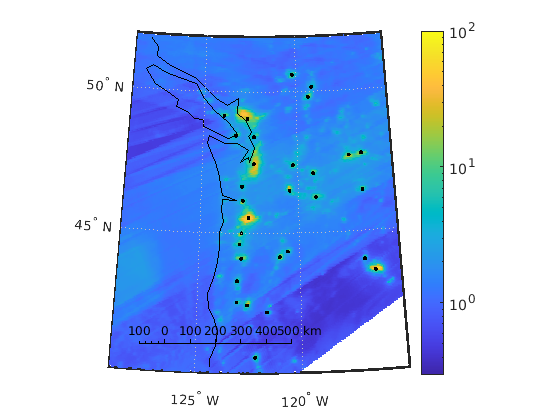}
    \caption{Cities along the North American west coast identified as connected areas with counts/GTU above 5, indicated here by black points. The largest areas seen here are Vancouver, Seattle and Portland. }
    \label{fig:example_city}
\end{figure} 

\begin{figure}[htb]
    \centering
    \includegraphics[width = 0.6\textwidth]{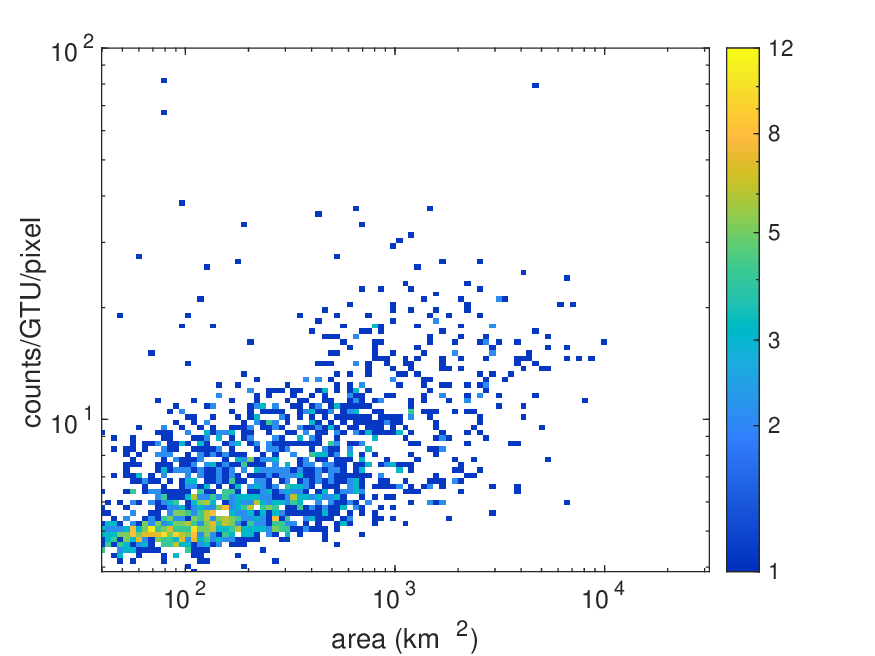}
    \caption{Histogram of the distribution of number of observations as a function of recorded counts/GTU/pixel (y-axis) of cities and their approximate area (x-axis). 
    The plot shows a gradual growth of the average luminosity with the size of the urban area.  }
    \label{fig:area_city_plot}
\end{figure}

\begin{figure}[H]
    \centering
    \includegraphics[width = .8
    \textwidth]{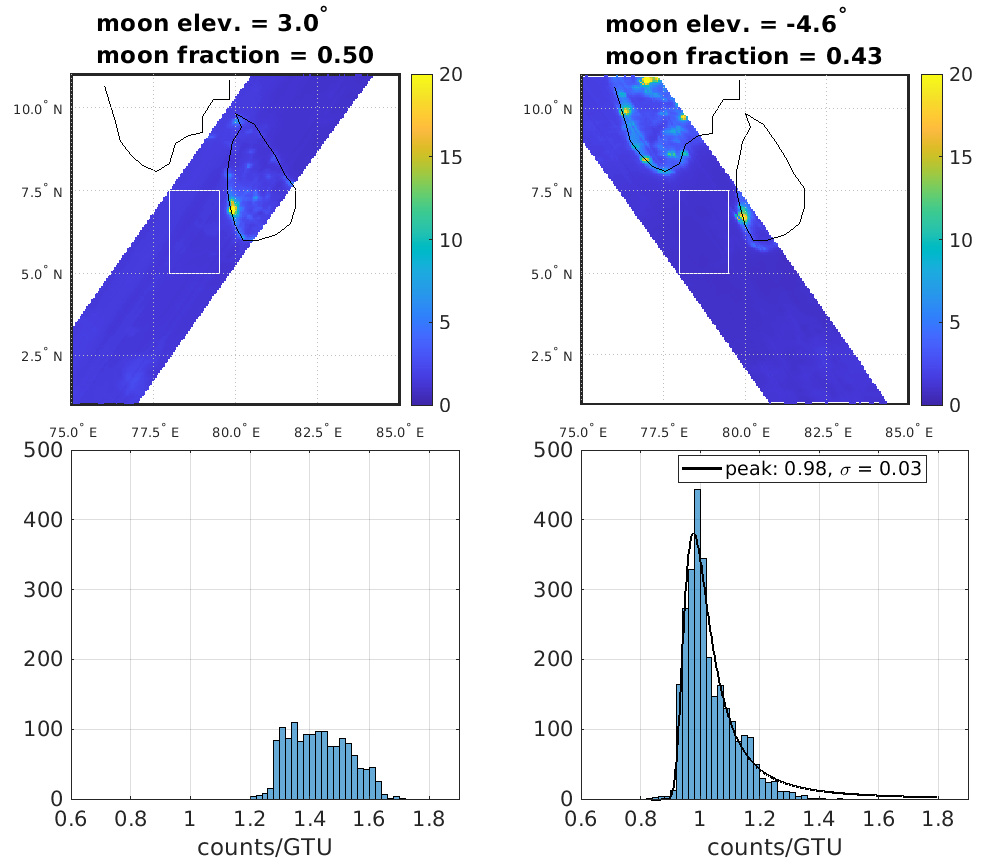}
    \caption{\textbf{Top: }UV maps of two spatially overlapping passages ($0.05^\circ\times0.05^\circ$ map cells) near the Sri-Lankan city of Colombo on 2020-03-02 (left) and 2020-03-31 (right). \textbf{Bottom:} Histograms of the area indicated by the rectangle, with a Landau distribution fit in black. }
    \label{fig:overlap1}
\end{figure}

\begin{figure}[H]
    \centering
    \includegraphics[width = .8
    \textwidth]{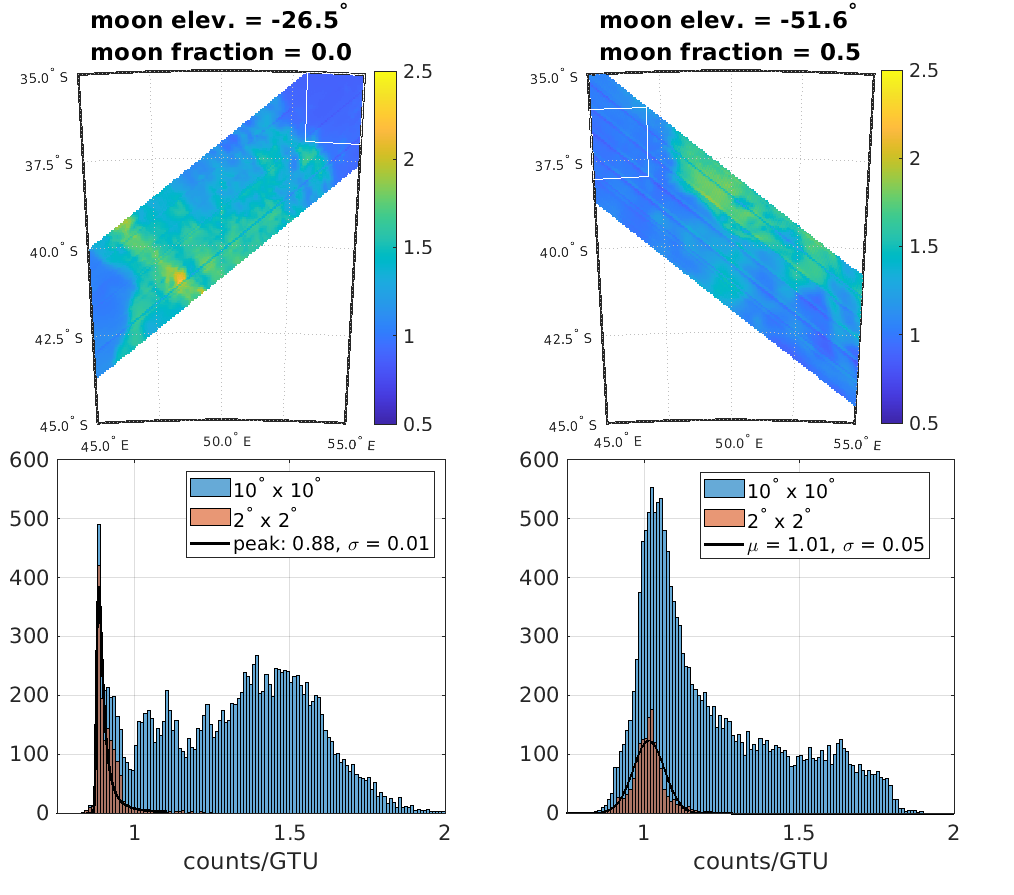}
    \caption{\textbf{Top: }UV maps of two spatially overlapping passages ($0.05^\circ\times0.05^\circ$ map cells) in the Indian Ocean, south of Madagascar on 2020-03-02 (left) and 2020-03-31 (right). \textbf{Bottom:} Histograms of the maps in the top panel, with a blue histograms representing the entire areas in and yellow histograms representing the area indicated by rectangles. The black lines show the corresponding Landau (left) and Gaussian (right) fits for the indicated areas.}
    \label{fig:overlap2}
\end{figure}

\textcolor{black}{Overlapping passages over the same ground location at different times, provide the opportunity to test the stability of the instrument and the reliability of the algorithms as a function of time. Figure~\ref{fig:europe_map} shows the similarity of the response of Mini-EUSO with time. In order to study long-term variations (due to natural or artificial conditions) we also have to evaluate  the effect of clouds and moonlight on the ground emissions. For instance, Figure~\ref{fig:overlap1} compares two regions off the western coast of Sri Lanka and shows that a Moon $3^\circ$ above the horizon can increase the average number of counts recorded over the same ocean area by $\simeq 40\%$.   
Another pair of  passages over the southern Indian Ocean is shown in Figure~\ref{fig:overlap2}. The Moon is below the horizon in both passages, but the presence of clouds on the overlapping region  results in a long tail of higher counts ($\simeq 1.5$~counts).   Indeed, if we restrict to two non-overlapping (but nearby) regions with less clouds, we have more similar distributions, although in the second case the average is slightly higher due to residual clouds. 
Future work - with a larger dataset - will involve the study of seasonal and long-term variation that we have to take into account. }

Because of a higher density of anthropogenic lights, such as cities, more areas with high counts/GTU are observed over land areas than sea. An example of such areas are shown in Figure~\ref{fig:example_city}. Here, patches of interconnected map cells with average counts/GTU above a threshold have been selected in a two-step approach. A first selection of areas with counts/GTU above 5 was made, with the area of each patch estimated by converting the number of $0.05^\circ\times0.05^\circ$ bins to km$^2$ based on the latitude in the centre of the area. To separate metropolitan areas of several cities, the same procedure was applied within areas larger than 1000~km$^2$, but with a threshold of 10 counts/GTU. For comparison, the city of Rome has an area of 1285~km$^2$ and a metropolitan area of 5686~km$^2$ (OECD, \url{http://measuringurban.oecd.org/}). The area around Rome identified using this method is 3617~km$^2$ using a threshold of 5 counts/GTU, and 1896~km$^2$ using 10~counts/GTU. In the corresponding cases, the average counts/GTU within the area is 12.0, and 15.0, respectively. 

Figure~\ref{fig:area_city_plot} shows, on a global scale, the relation between the average luminosity in counts/GTU/pixel (or counts/GTU/36~km$^2$) and the size of the urban  regions. 
The plot suggests that the average luminosity of a city tends to increase  with the increase of the area of the  city.

\section{UHECR observations}
 \label{sec:UHECR}
Beyond their immediate use in Earth observation, the above measurements can also be useful to evaluate the efficiency of  space-borne UHECR detectors and compare the results with the estimates coming from simulations (see e.g.~\citep{ADAMS201376}). 
\textcolor{black}{  Most of the dead time in UHECR observations from space comes from the requirement to be in local night, that is in the Earth's shadow. This period depends on the orbit of the instrument. In case of satellite-borne detectors, specific orbits can be devised to maximize the observation time (e.g. sun-synchronous polar orbits with the satellite on the night-day terminator, looking toward the night region of the Earth). For a generic low Earth orbit, a higher inclination means more sunlight and a lower efficiency. Furthermore, the day-night cycle depends on the beta angle, the angle  between the orbital plane of the satellite and the Sun-Earth vector. When is $\beta=90^\circ$ the ISS is always in sunlight and no operations are possible, whereas when $\beta=0^\circ$ the duration of the local night of the station is maximized. For instance, in Figure \ref{fig:solar_elev_example} we see that Mini-EUSO spends $\simeq 32$~minutes with the Sun below $30^\circ$: for a 90-minute orbit of the station, this translates in an efficiency of $\simeq 35\%$. This exposure is then reduced  by the presence of the Moon, clouds, lightning, towns and low latitude aurorae. }
In Figure~\ref{fig:hist_land_sea_cdf}, we see that the percentage of time where the background is $\leq$ 1 count/GTU in moonless, cloudless conditions is $\simeq$ 77\% over the sea  and $\simeq 63\% $ over the ground. Moonless nights, but with clouds present, exhibit a higher average  background, resulting in only 30\%  (sea) and 20\% (ground) percentage of time  spent below 1 count/GTU. The presence of the Moon increases the average background: however, it is possible to see that for Moon phases up to 40\% the average background remains below 2~counts/GTU $\simeq 90\%$ of the time; furthermore, Moon phases between 40\%  and 60\% have less than 2~counts/GTU approximately $30\%$ of the time. \textcolor{black}{  Furthermore, the acquisition system of Mini-EUSO allows up to four D1 triggers in a 5.2 seconds acquisition period. Therefore, in case of high lighting activity or with noisy photomultiplier pixels, this can cause some dead time to the acquisition of UHECR. For details on the trigger  system and the relative dead time see \citep{BATTISTI20222750}. On the overall, the efficiency for cosmic ray trigger during the first months of acquisition is about $8.5\%$,  in good agreement   from the calculation of $13\%$ in \citep{ADAMS201376}. The lower value of about 30\% in real conditions is due to a number of factors, mostly the aforementioned noise of some pixels that result in spurious triggers. Subsequent firmware updates have masked these pixels from the trigger, resulting in a lower dead time. Furthermore, the location of the instrument inside the ISS constraints its use only to specific days and periods: although planning of observations tries to optimize the observations, sometimes the sessions are performed during compromise periods, thus further reducing the efficiency of acquisitions. Future detectors located outside of the ISS (or on free-flyier satellites) will suffer less from this constraint.  }

\begin{figure}[h]
    \centering
    \includegraphics[width = .8
    \textwidth]{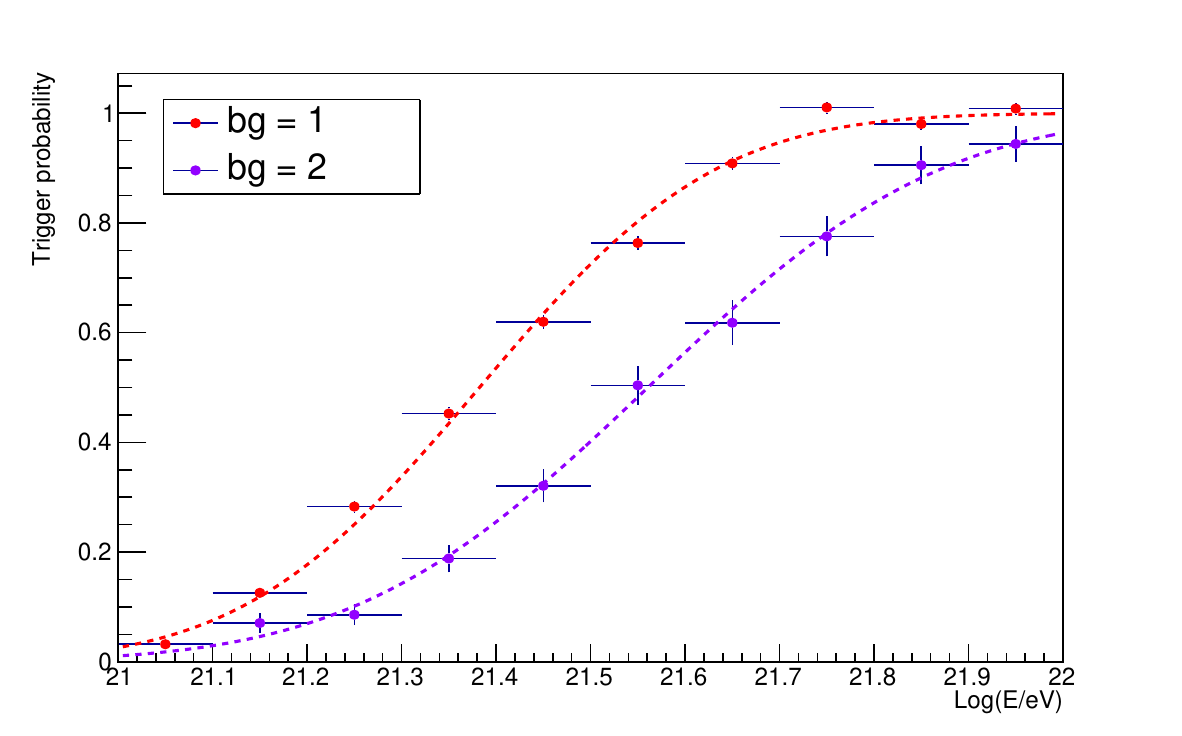}
    \caption{ UHECR efficiency curve as a function of energy for the Mini-EUSO detector in two different
background conditions (bg = 1 and bg = 2 counts/GTU) \citep{tesimarta}. Trigger threshold (the energy where trigger efficiency is $50\%$) is $10^{21.38}$~eV and  $10^{21.56} $~eV for  bg = 1~count/GTU and bg = 2~counts/GTU, respectively. }
    \label{fig:triggercurve}
\end{figure}

Knowing the average background conditions, we can use Montecarlo simulations of UHECR-generated EAS propagating in the atmosphere \citep{FENU2019ICRC...36..252F} to   estimate the detection efficiency as a function of the primary cosmic ray. Figure~\ref{fig:triggercurve} shows the efficiency curve as a function of the energy of the primary assuming a Poissonian distributed background of 1 and 2~counts/GTU. 
The higher the energy of the primary, the brighter the EAS and the more likely it is to  trigger the event \citep{Mini-EUSO_trigger}. The efficiency curve can be fitted with a sigmoidal error function \citep{auger_error}: 

\begin{equation}
\epsilon(E) = \frac{1}{2} \left[ 1 + erf \left( \frac{\log_{10} (E/{\rm eV}) - p_0}{p_1} \right) \right]
\end{equation}
where $p_0$ defines the efficiency threshold (trigger efficiency of 50\%) and $p_1$ the steepness of the slope.
The fit results give   $p_0 = 21.38\pm 0.03$  for a background of 1~counts/GTU, and  $p_0 = 21.56\pm    0.09$  for a background of 2 counts/GTU.
Under these conditions, we see that the detection threshold (assuming a triggering efficiency of $\simeq 50\% $) of  Mini-EUSO is  $10^{21.38}$ and $10^{21.56}$~eV for backgrounds of 1 and 2~counts/GTU, respectively \citep{tesimarta}.

The high threshold of Mini-EUSO is due to the size of the optics, which were constrained by the window of the ISS. With appropriate scaling, this result can be applied to similar detectors, and indeed the energy threshold of UHECR fluorescence telescopes is mostly determined by their lens (or mirror) size. The energy threshold also grows with increasing background (and thus the pixel field of view).  Since the highest energy cosmic ray event detected so far has an energy of $3.2\cdot 10^{20}$~eV \citep{1995ApJ...441..144B}, a  search for UHECR at these energies  will result in a lower limit on the  particle flux, thus requiring future detectors with larger optics, such as K-EUSO \citep{k-euso-universe} or POEMMA \citep{2019BAAS...51g..99O, Olinto_2021}, to achieve detection of UHECRs from space.  
In the planned K-EUSO detector, the lens size is 3~m$^2$, with an  estimated  energy threshold of $10^{19.2}$~eV \citep{k-euso-universe}.   Since the lens size is $\simeq 95$ times larger than Mini-EUSO and the pixel field of view is  $\simeq 1/110$ times smaller than Mini-EUSO ($0.6 \times 0.6$~km$^2$), the average  background light  in Mini-EUSO is thus comparable to that expected in  K-EUSO and similar larger  space-borne detectors such as POEMMA.

\section{Conclusions}

In this work, we have described observations from space in the near UV range (290 - 430~nm) from the ISS with the Mini-EUSO telescope. This detector operates with high sensitivity (single photoelectron counting), high temporal resolution (from 2.5~$\mu$s), and high spatial resolution ($\simeq~6.3$~km) in a wavelength range which is rarely studied, and the systematic nature of its observations from space allow it to contribute to the study of various phenomena which take place on our planet. In the context of general Earth observation,  in this paper we have focused on emissions taking place on the timescale of $\simeq 1$~second, as measured by free-running (non-triggered) acquisitions every 40.96~ms. 

We have  accounted for the actual detector response using a bootstrapped flat-fielding technique, which, under a few basic assumptions, allows us to calibrate the instrument response from the observational data itself. We then characterized the impact of the Moon elevation on the total UV background and the constraints posed by the solar elevation. Airglow and Moonlight, reflected by the Earth's surface and clouds, acts not only as a background to emission sources present on Earth, but also as an illumination source for the study of non-emitting terrain and atmospheric phenomena. To account for the fact that observations of surface emissions and reflections are affected by cloud cover, we correlate our observations with external cloud coverage predictions. Doing so, we have found a good agreement between the predicted cloud cover and the actual observed conditions, and Mini-EUSO data containing cloud formations also show the good resolution of the telescope, with the internal structure of cloud formations clearly visible. Using this cloud coverage information, we have estimated the dependence of the UV background on the percentage cloud cover. 

We have also estimated the relationship between the observed counts and the absolute photon flux for point-like sources on the ground, diffuse ground sources, diffuse upper-atmospheric sources, and clouds. We estimate that Mini-EUSO is sensitive to any point-like $2.5~\mu$s timescale ground-level transient of power greater than $70\pm15$~W (at 395~nm).

We then produced night-time maps of the Earth in the UV by binning Mini-EUSO data into $0.1^{\circ}\times0.1^{\circ}$ cells over the Earth's surface. These maps show the clear impact of anthropogenic light sources on the night-time UV environment, such as those coming from towns and fishing.

Using such maps, we are able to segment contiguous urban areas, and have found that the apparent light density, in photons per unit time per km$^{2}$ of urban area increases with increasing size of the urban area. 

Finally, these observation provide an experimental check for simulated estimates of the observation conditions expected by future space-borne UHECR detectors and, therefore, their expected efficiency. 
In particular, we have characterized the airglow emissions of dark moonless nights over the sea and in different terrain conditions. This has allowed us to determine the percentage of time with a background below a few counts/GTU under various Moon and cloud conditions, and, appropriately scaled for aperture and pixel field of view, this information can be used to estimate the percentage time spent at a given EAS detection efficiency as a function of the primary cosmic ray energy for any space-borne UHECR detector.

Future work on this topic will involve the study of long-term variations in night-time UV emission at this timescale due to natural and artificial phenomena. As mentioned, the data presented here  are based   only on continuous acquisitions: triggered observations at faster time scales allow us to also study Transient Luminous Events, gravity waves, and meteors, search for hypothetical strange quark matter, and potentially observe UHECR events. 

The possibility of making the UV night-time maps of the Earth and associated data available on a public server is being discussed with the Italian and Russian Space Agencies. 

\section{Acknowledgements}

This work was supported  by the Italian Space Agency through the agreement n.~2020-26-Hh.0,  by the French space agency CNES, and by the National Science Centre in Poland grants 2017/27/B/ST9/02162 and 2020/37/B/ST9/01821.
This research has been supported by the Interdisciplinary Scientific and Educational School of Moscow University ``Fundamental and Applied Space Research'' and by  Russian State Space Corporation Roscosmos. The article has been prepared based on research materials collected in the space experiment ``UV atmosphere''. We thank the Altea-Lidal collaboration for providing the orbital data of the ISS. 

\bibliographystyle{aasjournal}
\bibliography{bibliography_MiniEUSO}

\end{document}